\pdfoutput=1 
\documentclass[useAMS,usenatbib,letterpaper]{mn2e}
\usepackage{amssymb}
\usepackage{amsmath}
\usepackage{multirow}
\usepackage{graphicx}
\usepackage[british]{babel} 
\usepackage[pdftex]{hyperref}
\usepackage{url}
\usepackage{color}
\usepackage{pdflscape}

\newcommand{\dif}{{\mathrm d}}
\newcommand{\ie}{\textit{i.e.}}
\def\fnu{\mathfrak{f}_\nu}

\def\lcdm{$\Lambda$CDM}

\begin{document}
  

\title[Constraints from clustering ratio with neutrinos]{Cosmological constraints from galaxy clustering in the presence of massive neutrinos}
\author[M. Zennaro, J. Bel, J. Dossett, C. Carbone, L. Guzzo] 
{M. Zennaro$^{1,3}$, J. Bel$^{2,3}$, J. Dossett$^{3}$,  C. Carbone$^{1,3,4}$, L. Guzzo$^{1,3}$\\
\\
$^{1}$Universit\`a degli studi di Milano - Dipartimento di Fisica, via Celoria, 16, 20133 Milano, Italy\\
$^2$Aix Marseille Univ, Universit\'e de Toulon, CNRS, CPT, Marseille, France\\
$^3$INAF - Osservatorio Astronomico di Brera, Via Brera, 28, 20121 Milano, Italy\\
$^4$INFN - Sezione di Milano, via Celoria, 16, 20133 Milano, Italy
} 

\date{Received date / Accepted date}

\maketitle


\begin{abstract}
The clustering ratio is defined as the ratio between the correlation function and the variance of the smoothed overdensity field. In $\Lambda$CDM cosmologies not accounting for massive neutrinos, it has already been proved to be independent from bias and redshift space distortions on a range of linear scales. It therefore allows for a direct comparison of measurements (from galaxies in redshift space) to predictions (for matter in real space). In this paper we first extend the applicability of such properties of the clustering ratio to cosmologies that include massive neutrinos, by performing tests against simulated data. We then investigate the constraining power of the clustering ratio when cosmological parameters such as the total neutrino mass $M_\nu$ and the equation of state of dark energy $w$ are left free. We analyse the joint posterior distribution of the parameters that must satisfy, at the same time, the measurements of the galaxy clustering ratio in the SDSS DR12, and the angular power spectrum of temperature and polarization anisotropies of the CMB measured by the Planck satellite. We find the clustering ratio to be very sensitive to the CDM density parameter, but not very much so to the total neutrino mass. Lastly, we forecast the constraining power the clustering ratio will achieve with forthcoming surveys, predicting the amplitude of its errors in a Euclid-like galaxy survey. In this case, we find it is expected to improve the constraint at 95\% level on the CDM density by 40\% and on the total neutrino mass by 14\%.
\end{abstract}
 
\begin{keywords}
bias, clustering, neutrinos
\end{keywords}


\section{Introduction}\label{sec:introduction}
Present-time as well as forthcoming galaxy surveys, while on the one hand will allow us to reach unprecedented precision on the measurement of the galaxy clustering in the universe, on the other hand will challenge us to produce more accurate and reliable predictions. The effect of massive neutrinos on the clustering properties of galaxies, that in the past has been either neglected or considered as a nuisance parameter, is nowadays regarded as one of the key points to be included in the cosmological model in order for it to reach the required accuracy. At the same time, while allowing for more realistic predictions of cosmological observables, this process also helps in shedding light on some open issues of fundamental physics, such as the neutrino total mass or the hierarchy of their mass splitting.

From the experiments measuring neutrino flavour oscillations, particle physics has been able to draw a constraint on the mass splitting of the massive eigenstates of neutrinos, and set a lower bound to the total neutrino mass, $M_\nu=\sum m_{\nu,i} \gtrsim 0.06$ eV at 95\% level \citep{Gonzalez-GarciaEtal2012,Gonzalez-GarciaMaltoniSchwetz2014,ForeroTortolaValle2014,EstebanEtal2017}.

On the other hand, the absolute scale of magnitude of neutrino masses is still an open issue. Beta decay experiments such as the ones carried out in Mainz and Troitsk have set as an upper limit at 95\% level on the electron neutrino mass of $m(\nu_e) < 2.2$ eV \citep{KrausEtal2005}. While future experiments like Katrin prospect much higher sensitivities, of order $0.2$ eV \citep{BonnEtal2011}, present day cosmology can already intervene in the debate about neutrino mass.

Since neutrinos are light and weakly interacting, they decouple from the background when still relativistic. Therefore, even at late times they are characterised by large random velocities that prevent them from clustering on small scales. As a consequence, neutrinos introduce a characteristic scale-dependent and redshift-dependent suppression of the clustering, whose amplitude depends on the value of their mass. In fact, the presence of massive neutrinos influences the evolution of matter overdensities in the universe, depending on their mass. Last years have seen a large number of works extensively studying the interplay between cosmology and neutrino physics \citep[see, for instance,][and references therein]{LesgourguesPastor2006,LesgourguesPastor2012,LesgourguesPastor2014}.

As we aim to describe the clustering of galaxies in cosmologies with massive neutrinos, we have to cope with the description of the galaxy-matter bias. As a matter of fact, galaxies do not directly probe the matter distribution in the universe, being in fact a discrete sampling of its highest density peaks. We choose to describe galaxy clustering through a recently introduced observable that, on sufficiently large scales, does not depend on the galaxy-matter bias, the clustering ratio \citep{BelMarinoni2014}.

In standard \lcdm{} cosmologies, this observable has already been proved to be a reliable cosmological probe for constraining cosmological parameters, being particularly sensitive to the amount of matter in the universe, as shown in \citet{BelMarinoniGranett2014}.

In this work we aim at studying how the clustering properties of galaxies are modified by the presence of neutrinos, and in particular we want to extend the clustering ratio approach to cosmologies including massive neutrinos. By proving that this observable maintains its properties, we want to exploit it to constrain the total neutrino mass. 

This paper will be organized as follows.
In Sec. \ref{sec::cr} we will introduce the statistical observable we are going to use, the clustering ratio, and its properties. We will show why this observable can be considered unaffected either by the galaxy-matter bias on linear scales and redshift-space distortions, and we will introduce its estimators.

In Sec. \ref{sec::nu} we will describe the effects of massive neutrinos on the matter and galaxy clustering. We will introduce the DEMNUni simulations, the set of cosmological simulations we use to test the properties of the clustering ratio in a cosmology with massive neutrinos. Finally we will show that the properties of the clustering ratio hold as well in cosmologies that include massive neutrinos, in particular confirming the independence of the clustering ratio from bias and redshift-space distortions on linear scales in the DEMNUni simulations.

Sec. \ref{sec::res} is devoted to presenting our results. We use a measurements of the clustering ratio in the Sloan Digital Sky Survey Data Release 7 and 12 to draw a constraint on the total neutrino mass and on the equation of state of dark energy. In particular we study the joint posterior distribution of the parameters of the model, including $M_\nu$ and $w$, obtained from the clustering ratio measurement and the latest cosmic microwave temperature and polarization anisotropy data from the Planck satellite.

\section{Clustering ratio}\label{sec::cr}

In order to describe the statistical properties of the matter distribution in the universe, we introduce the overdensity field

\begin{equation}
\delta(\boldsymbol{x},t) = \dfrac{\rho(\boldsymbol{x},t)}{\bar{\rho}(t)} - 1,
\end{equation}
where $\rho(\boldsymbol{x},t)$ is the value of the matter density at each spatial position, while $\bar{\rho}(t)$ represents the mean density of the universe.

This is assumed to be a random field with null mean. Information on the distribution must therefore be sought in its higher order statistics, such as the variance $\sigma^2=\langle \delta^2 (\boldsymbol{x}) \rangle_c$ and the 2-point autocorrelation function $\xi (r) = \langle \delta (\boldsymbol{x}) \delta (\boldsymbol{x} + \boldsymbol{r}) \rangle_c$ of the field. Here $\langle \cdot \rangle_c$ denotes the cumulant moment, or connected expectation value \citep{Fry1984}.

In this work we will always consider the matter distribution smoothed on a certain scale $R$ by evaluating the density contrast in spherical cells, \ie{}

\begin{equation}
\delta_R (\boldsymbol{x}) = \int \! \delta (\boldsymbol{x}^\prime) W \left(\frac{|\boldsymbol{x}-\boldsymbol{x}^\prime|}{R}\right) \dif^3 \boldsymbol{x}^\prime,
\end{equation}
where $W$ is the spherical top-hat window function. As a consequence, the variance and correlation function will be smoothed on the same scale, and will be denoted $\sigma^2_R$ and $\xi_R (r)$.

An equivalent description of the statistical properties of the matter field can be obtained in Fourier space in terms of the matter power spectrum. Starting from the Fourier transform of the matter overdensity field, 

\begin{equation}
\hat{\delta}(\boldsymbol{k}) = \int \! \dfrac{\dif^3 \boldsymbol{x}}{(2 \pi)^3} e^{ -i \boldsymbol{k} \cdot \boldsymbol{x} } ~ \delta(\boldsymbol{x})~,
\end{equation}
the matter density power spectrum is defined according to
\begin{equation}
\langle \hat{\delta}(\boldsymbol{k}_1) \hat{\delta}(\boldsymbol{k}_2) \rangle = \delta ^D (\boldsymbol{k}_1 + \boldsymbol{k}_2) P(\boldsymbol{k}_1) ~ ,
\end{equation}
while the adimensional power spectrum can be written as $\Delta^2(k) = 4 \pi P(k) k^3$. The variance and correlation function of the matter field are linked to its power spectrum, representing in fact different ways of filtering it. The variance is the integral over all the modes, modulated by the Fourier transform of the filtering function $\hat{W}$,
\begin{equation}
\sigma^2_R = \int_0^\infty \! \Delta^2(k) \hat{W}^2(kR) ~ \dif \ln k,
\end{equation}
and the correlation function is, in addition, modulated by the zero-th order spherical Bessel function $j_0 (x) = \sin(x)/x$,

\begin{equation}
\xi_R(r) = \int_0^\infty \! \Delta^2(k) \hat{W}^2 (kR) j_0(kr) ~ \dif \ln k.
\end{equation}
The explicit expression of $\hat{W}(kR)$ is
\begin{equation}
\hat{W}(kR) = \frac{3}{kR} j_1(kR) = 3 \frac{\sin(kR) - kR \cos (kR)}{(kR)^3}.
\end{equation}
However, in practice, we are not able to directly access the matter power spectrum. The reason is that the galaxies we observe do not directly probe the distribution of matter in the universe. In fact, they represent a discrete biased sampling of the underlying matter density field and the biasing function is, a priori, not known. A way to overcome this problem is to refine the independent measurements of the bias function through weak lensing surveys. Otherwise, one can parametrise the bias adding  additional nuisance parameters to the model and, consequently, marginalize over them. 

On the other hand, a completely different approach is to seek new statistical observables, which can be considered to be unbiased by construction. This is the path followed by \citet{BelMarinoni2014} by introducing the clustering ratio,

\begin{equation}
\eta_R (r) \equiv \dfrac{\xi_R(r)}{\sigma_R^2},
\label{eq::cr-def}
\end{equation}
that is the ratio of the correlation function over the variance of the smoothed field.

We assume that the relation between the matter density contrast and the galaxy (or halo) density field is a local and deterministic mapping, which is enough regular to allow a Taylor expansion \citep{FryGaztanaga1993} as

\begin{equation}
\delta_{g,R} = F(\delta_{R}) \simeq \sum_{i = 1}^N \dfrac{b_i}{i!} \delta_R^i.
\label{bias}
\end{equation}
Moreover we assume the growth of fluctuations to occur hierarchically \citep[see][]{BernardeauEtal2002, BelMarinoni2012}, so that each higher order cumulant moment can be expressed according to powers of the variance and 2-point correlation funtion,
\begin{equation}
\begin{split}
\langle \delta_{R}^n \rangle_c &= S_{n} \sigma_R^{2(n-1)}~,\\
\langle \delta_{i,R}^n  \delta_{j,R}^m \rangle_c &= C_{nm} \xi_{R}(r) \sigma_R^{2(n+m-2)}~,
\end{split}
\end{equation}
the former applying to the 1-point statistics and the latter to the 2-point ones.

It has been shown by  \citet{BelMarinoni2012} that the bias function only modifies the clustering ratio of galaxies at the next order beyond the leading  one and that it is not sensitive to third order bias
\begin{eqnarray}
\eta_{g,R} (r) &\simeq & \eta_R(r) + \dfrac{1}{2} c_2^2 \eta_R(r) \xi_R(r) \nonumber\\
& &- \left\{ ( S_{3,R} - C_{12,R}) c_2 + \dfrac{1}{2} c_2^2 \right\} \xi_R(r)  \label{eq::cr-bias}
\end{eqnarray}
where $c_2 \equiv b_2/b_1$. 
By choosing a sufficiently large smoothing scale, the higher order contribution in Eq. (\ref{eq::cr-bias}) becomes negligible and we obtain 
\begin{equation}
\eta_{g,R} (r) = \eta_{R} (r),
\label{eq:CR-identity}
\end{equation}
meaning that in this case the clustering ratio of galaxies can be directly compared to the clustering ratio predicted for the matter distribution. 

The local biasing model is not the best way of describing the bias function between matter and haloes/galaxies \citep{MoWhite1996,ShethLemson1999,SomervilleEtal2001,Casas-MirandaEtal2002}. It can be improved by introducing a non-local component depending on the tidal field.  However, it has been shown by \cite{ChanScoccimarroSheth2012} and by \cite{BelHoffmannGaztanaga2015} that, when dealing with statistical quantities which are averaged over all possible orientations, then the non-local component is degenerate with the second order bias $c_2$, thus expression (\ref{eq::cr-bias}) remains valid and we do not consider non-local bias in our analysis.

On linear scales, the clustering ratio is expected to be independent from redshift. Since, in a $\Lambda$CDM universe, we can write (normalizing all quantities to the present time variance on the scale $r_8 = 8 ~ h^{-1}$ Mpc, $\sigma_8^2(z=0)$) the evolution of the variance and the correlation function as
\begin{equation}
\begin{split}
\sigma^2_R (z) &= \sigma^2_8 (z=0) D^2(z) \mathcal{F}_R\\
\xi_R (r,z) &= \sigma^2_8 (z=0) D^2(z) \mathcal{G}_R (r) ,
\end{split}
\end{equation} 
where $D(z)$ is the linear growth factor of matter density fluctuations and
\begin{displaymath}
\begin{split}
\mathcal{F}_R &= \dfrac{\int_0^\infty \! \Delta ^2(k) \hat{W}^2 (kR) \dif \ln k}{\int_0^\infty \! \Delta ^2(k) \hat{W}^2 (kr_8) \dif \ln k} \\
\mathcal{G}_R (r) &= \dfrac{\int_0^\infty \! \Delta ^2(k) \hat{W}^2 (kR) j_0 (kr) \dif \ln k}{\int_0^\infty \! \Delta ^2(k) \hat{W}^2 (kr_8) \dif \ln k} ,
\end{split}
\end{displaymath}
depend only on the shape of the power spectrum. Hence, the clustering ratio $\xi_R(r)/\sigma^2_R = \mathcal{G}_R (r)/\mathcal{F}_R$, does not depend on $D(z)$, which cancels outs. 

In practice, we include weak nonlinearities which introduce a small, but nevertheless detectable, redshift dependence.

Not only measures of the clustering of galaxies are biased with respect to predictions for the matter field, but they also are affected by the peculiar motion of galaxies. This motion introduces a spurious  velocity component (along the line-of-sight) that distorts the redshift assigned to galaxies. Since, for the clustering ratio, we are interested in large smoothing scales and separations, we can focus on the linear scales, where the only effect is due to the coherent motion of infall of galaxies towards the overdense regions in the universe.

We can link the position of a galaxy (or dark matter halo) in real space to its apparent position in redshift-space. Let us denote $\boldsymbol{r}$ as the true comoving distance along the line-of-sight; in redshift space it becomes

\begin{equation}
\boldsymbol{s} = \boldsymbol{r} + \dfrac{v_{p \parallel}~(1+z)}{H(z)} \hat{\boldsymbol{e}}_{\boldsymbol{r}},
\label{rtos}
\end{equation}
where $v_{p \parallel}$ is the line-of-sight component of the peculiar velocity and $\hat{\boldsymbol{e}}_{\boldsymbol{r}}$ is the line-of-sight versor. Considering the Fourier space decomposition of the density contrast, the relation linking its value in redshift space to the one in real space \citep{Kaiser1987} is
\begin{equation}
\delta^s(\boldsymbol{k}) = (1 + f \mu^2) \delta(\boldsymbol{k}),
\end{equation}
where quantities in redshift space are expressed with the superscript \textit{s} and $\mu$ is the cosine of the angle between the wavemode $\boldsymbol{k}$ and the line-of-sight. Here $f$ is the so called growth rate, defined as the logarithmic derivative of the growth factor of structures with respect to the scale factor, $f \equiv \dif \ln D / \dif \ln a$. Averaging over all angles $\vartheta$, the variance and the correlation function in redshift space result modified by the same multiplicative factor

\begin{equation}
\begin{split}
\sigma^{s\,2}_R &= K \sigma^2_R\\
\xi_R^s(r) &= K \xi_R (r)
\end{split}
\end{equation}
where $K = 1 + 2f/3 + f^2/5$ is the Kaiser factor. As a consequence, the clustering ratio is unaffected by redshift-space distortions on linear scales. This argument allows us to rewrite the identity (\ref{eq:CR-identity}) as

\begin{equation}
\eta^s_{g,R} (r) \equiv \eta_{R} (r),
\label{eq:CR-identity-zspace}
\end{equation}
meaning that, by properly choosing the smoothing scale $R$ and the correlation length $r$, measures of the clustering ratio from galaxies in redshift space can be directly compared to predictions for the clustering ratio of matter in real space.

\subsection{Estimators}\label{subsec:estimators}

The clustering ratio can be estimated from count-in-cells, where, under the assumption of ergodicity, all ensemble averages become spatial averages. We follow the counting process set up  by \citet{BelMarinoni2012}, we define the discrete density contrast as

\begin{equation}
\delta_{N,i} = \frac{N_i}{\bar{N}^{\phantom{|}}} -1 ,
\end{equation}
where $N_i$ is the number of objects in the \textit{i}-th cell and $\bar{N}$ is the mean number of objects per cell. The estimator of the variance is therefore
\begin{equation}
\hat{\sigma}_R^2 = \frac{1}{p} \sum_{i=1} ^p \delta_i ^2
\end{equation}
and the one of the correlation function is
\begin{equation}
\hat{\xi}_R (r) = \frac{1}{pq} \sum_{i=1} ^p \sum_{j=1} ^q \delta_i \delta_j
\end{equation}
leading to the definition of the estimator of the clustering ratio as
\begin{equation}
\hat{\eta}_R (r) = \frac{\hat{\xi}_R (r)}{\hat{\sigma}_R^2}.
\end{equation}
Throughout this work we will often express the correlation length $r$ as a multiple of the smoothing scale, \ie{} $r=n~R$.

Since we are dealing with a discrete counting process, the shot noise needs to be properly accounted for. We follow the approach of \cite{BelMarinoni2012} and correct the estimator of the variance according to
\begin{equation}
\hat{\sigma}_R^2 = \langle \delta_n^2(\boldsymbol{x}) \rangle - \frac{1}{\bar{N}} = \frac{1}{p} \sum_{i=1} ^p \delta_i ^2 - \frac{1}{\bar{N}},
\end{equation}
where $\bar{N}$ is the mean number of objects per cell. On the other hand, the correlation function needs no correction, as long as the spheres do not overlap. 

\subsection{Effects of massive neutrinos}\label{subsec:neutrinos}

We introduce massive neutrinos as a subdominant dark matter component. For simplicity, we consider three degenerate massive neutrinos, with total mass $M_\nu = \sum_i m_{\nu,i}$ and  present-day neutrino energy density in units of the critical density of the universe $\Omega_{\nu,0}h^2 = M_\nu/(93.14 ~ \mathrm{eV})$. The neutrino fraction is usually expressed with respect to the total matter as $\fnu=\Omega_\nu/\Omega_m$. For a more complete treatment of neutrinos in cosmology, we refer the reader to \cite{LesgourguesPastor2006,LesgourguesPastor2012,LesgourguesPastor2014}.

Neutrinos of sub-eV mass, which seem to be the most likely candidates both from particle physics experiments and cosmology, decouple from the primeval plasma when the weak interaction rate drops below the expansion rate of the universe, at a time when the background temperature is around $T\simeq 1~{\rm MeV}$. This corresponds to a redshift $1+z_{\rm dec} \sim 10^9$. Since the redshift of their non-relativistic transition, obtained equating their rest-mass energy and their thermal energy, is given by
\begin{equation}
1+z_{nr}\simeq 1890 ~ \dfrac{m_{\nu,i}}{1~\mathrm{eV}},
\end{equation} 
when neutrinos decouple, they are still relativistic. As a consequence, since the momentum distribution of any species is frozen at the time of decoupling, neutrino momenta keep following a Fermi-Dirac distribution even after their non-relativistic transition, and neutrinos end up being characterised by a large velocity dispersion.
An effective description of the evolution of neutrinos can be achieved employing a fluid approximation \citep{ShojiKomatsu2010}. In this framework we can define a neutrino pressure, $p_\nu = w_\nu \rho_\nu c^2$, computed integrating the momentum distribution. Such pressure is characterised by an effective adiabatic speed of sound \citep{BlasEtal2014}
\begin{equation}
c_{s,i} = 134.423 ~(1+z)~ \dfrac{1{\rm eV}}{m_{\nu,i}} ~ \mathrm{km}~\mathrm{s}^{-1},
\end{equation}
that represents the speed of propagation of neutrino density perturbations. Such speed of sound defines the minimum scale under which neutrino perturbations cannot grow, called the free streaming scale. It corresponds to a wavenumber 

\begin{equation}
k_{FS} (z) = \left[\dfrac{4\pi G \bar{\rho} a^2}{c_s^2}\right]^{1/2} = \left[ \dfrac{3}{2} \dfrac{H^2 \Omega_m(z)}{(1+z)^2 c_s^2} \right]^{1/2}, 
\end{equation}
or a proper wavelength 
\begin{equation}
\lambda_{FS} = 2\pi a/k_{FS}. 
\end{equation}
At each redshift, neutrino density fluctuations of wavelength smaller than the free streaming scale are suppressed, their gravitational collapse being contrasted by the fluid pressure support. As a consequence, neutrinos do not cluster on small scales and remain more diffuse compared to the cold matter component.

Neutrino free streaming does not only affect the evolution of neutrino perturbations, in fact it affects the evolution of all matter density fluctuations. We can model the growth of matter fluctuations employing a two-fluid approach \citep{BlasEtal2014,ZennaroEtal2016}. In this case, the solution of the equations of growth for the neutrino and cold matter fluids are coupled,
\begin{equation}
\begin{cases}
\ddot{\delta}_{cb} + \mathcal{H}\dot{\delta}_{cb} - \dfrac{3}{2}\mathcal{H}^2\Omega_m \left\{ \fnu \delta_\nu +(1-\fnu)\delta_{cb} \right\} = 0\\
\ddot{\delta}_{\nu} + \mathcal{H}\dot{\delta}_{\nu} - \dfrac{3}{2}\mathcal{H}^2\Omega_m \left\{ \left[ \fnu - \dfrac{k^2}{k_{FS}^2} \right] \delta_\nu + (1-\fnu) \delta_{cb} \right\} = 0,
\end{cases}
\label{eq:coupled-growth}
\end{equation}
where derivatives are taken with respect to conformal time, $\dif \tau = \dif t/a$, and both the Hubble function and the matter density parameter are functions of time, $\mathcal{H}=\mathcal{H}(\tau)$ and $\Omega_m=\Omega_m(\tau)$.

The coupling of these equations requires the evolution of the CDM density contrast to be scale dependent, unlike in standard $\Lambda$CDM cosmologies. We therefore expect to find an observable suppression even in the CDM+baryon power spectrum, starting from the mode corresponding to the size of the free-streaming scale at the time of the neutrino non-relativistic transition, $k_{nr} = k_{FS}(z_{nr})$, and affecting all the scales smaller than this one.

\section{Clustering Ratio with massive neutrinos}\label{sec::nu}
In order to investigate the behaviour of the clustering ratio in cosmologies with massive neutrinos, \ie{} whether it maintains all the properties described in Sec. \ref{sec::cr}, we analyse the cosmological simulations ``Dark Energy and Massive Neutrino Universe'' (DEMNUni), presented in \citet{CastorinaEtal2015} and \citet{CarbonePetkovaDolag2016}. 

These simulations have been performed using the \verb+Gadget-III+ code by \cite{VielHaehneltSpringel2010} based on the \verb+Gadget+ simulation suite \citep{SpringelEtal2001,Springel2005}. This version includes three active neutrinos as an additional particle species\footnote{The simulations do not account for an effective neutrino number $N_{\rm{eff}} > 3$, as possible neutrino isocurvature perturbations which could produce larger $N_{\rm{eff}}$ \citep[therefore affecting galaxy and CMB statistics][]{CarboneMangilliVerde2011} are currently excluded by present data \citep[see, eg,][]{DiValentinoMelchiorri2014}}.

The DEMNUni project comprises two set of simulations. The first one, which is the one considered in the present work, includes 4 simulations, each implementing a different neutrino mass. Besides the reference $\Lambda$CDM simulation, which has $M_\nu=0$ eV, the other ones are characterised by $M_\nu = \lbrace 0.17, 0.30, 0.53 \rbrace$ eV. The second set includes 10 simulations, exploring different combinations of neutrino masses and dynamical dark energy parameters. 

\begin{table}
\centering
\begin{tabular}{lrrrr}
\hline
&$\Lambda$CDM&NU0.17&NU0.30&NU0.53\\
\hline
$M_\nu~[{\rm eV}]$ & 0 & 0.17 & 0.30 & 0.53 \\
$\Omega_c$ & 0.27 & 0.2659 & 0.2628 & 0.2573 \\
$\sigma_{8,cc}$ & 0.846 & 0.813 & 0.786 & 0.740 \\
\hline
\end{tabular}
\caption{The cosmological parameters that vary among the 4 DEMNUni simulations considered in the present work, depending on the assumed neutrino total mass.}
\label{tab::demnuni}
\end{table}

All simulations share the same Planck-like cosmology, with Hubble parameter $H_0 = 67 ~ km ~ s^{-1} ~ \mathrm{Mpc}^{-1}$, baryon density parameter $\Omega_b = 0.05$, primordial spectral index $n_s = 0.96$, primordial amplitude of scalar perturbations $A_s = 2.1265 \times 10^9$ (at a pivotal scale $k_p = 0.05 ~ \mathrm{Mpc}^{-1}$) and optical depth at the time of recombination $\tau = 0.0925$. The density parameter of the cold dark matter, $\Omega_{cdm}$, is adjusted in each simulation, depending on the neutrino mass, so that all simulations share the same total matter density parameter $\Omega_m = 0.32$, see Tab. \ref{tab::demnuni}. Each simulation follows the evolution of $2048^3$ CDM particles and, when present, $2048^3$ neutrino particles, in a comoving cube of $2~h^{-1}$ Gpc side. The mass of the CDM particle is $\sim 8\times 10^{10} h^{-1} M_{\odot}$, and changes slightly depending on the value of $\Omega_{cdm}$. All simulations start at an initial redshift $z_{in}=99$ and reach $z=0$ with 62 comoving outputs at different redshifts. In this work we focus on the snapshots at redshift $z=0.48551$ and $z=1.05352$.

Dark matter haloes have been identified through a Friend-of-Friends (FoF) algorithm with linking length $b=0.2$ and setting the minimum number of particles needed to form a halo to $32$. Thus, the least massive haloes have mass of about $2.6\times 10^{12} h^{-1} M_{\odot}$. In order to check the stability of our results regarding the choice of the definition of a halo we also have access to halo catalogues where haloes have been identified using spherical over-densities.
For the purpose of the present work we constructed halo catalogues in redshift space by modifying the positions along the $z$ direction according to the projected velocity (properly converted in length) in that direction (see Eq. \ref{rtos}). 

Regarding error estimation, being these simulations very large, a jackknife method has been implemented by subdividing the box in 64 sub-cubes. The standard error on the measured value of $\eta_R (r)$ is then taken to be the dispersion obtained from the jackknife process
\begin{equation}
\sigma_{\eta_R}^2 = \frac{N_j - 1}{N_j} \sum_{i = 1}^{N_j} \left[ \eta_{R,i} (r) - \bar{\eta}_R (r) \right]^2,
\end{equation} 
where $N_j$ is the number of jackknife resamplings, in our case $N_j = 64$.

In the following, we first check the reliability of the clustering ratio in the presence of massive neutrinos. In particular we are interested in proving that the identity $\eta_{R,g}^z (r) \equiv \eta_R (r)$ still holds. To this end, we must prove that the clustering ratio at the scales of interest does not depend on the galaxy-matter bias (so that we can compare predictions for matter and measures from galaxies) and that it is not affected by redshift-space distortions (to be safe when comparing the real space predictions to measures obtained in a galaxy redshift survey).

\subsection{Bias sensitivity}

In order to test the independence of the clustering ratio from the bias on linear scales, we divide the dark matter haloes in $9$ mass bins, reported in Tab. \ref{tab:mass-bins}. The various halo populations evolve in  a different way, therefore they present different biasing functions with respect to the dark matter field. Thus, we will use the linear bias $b_L$ to characterise each halo sample. 
In Tab. \ref{tab:pop-bins} we show how both the FoFs and the spherical overdensities from the simulations populate these mass bins in the simulations. Due to the minimum number of particle required to identify a halo, the first two mass bins do not contain any. On the other hand, the only mass limit to the spherical over-densities is given by the mass resolution of the simulation, hence all the mass bins are populated.

\begin{figure*}
\centering
\includegraphics[width=0.49\textwidth]{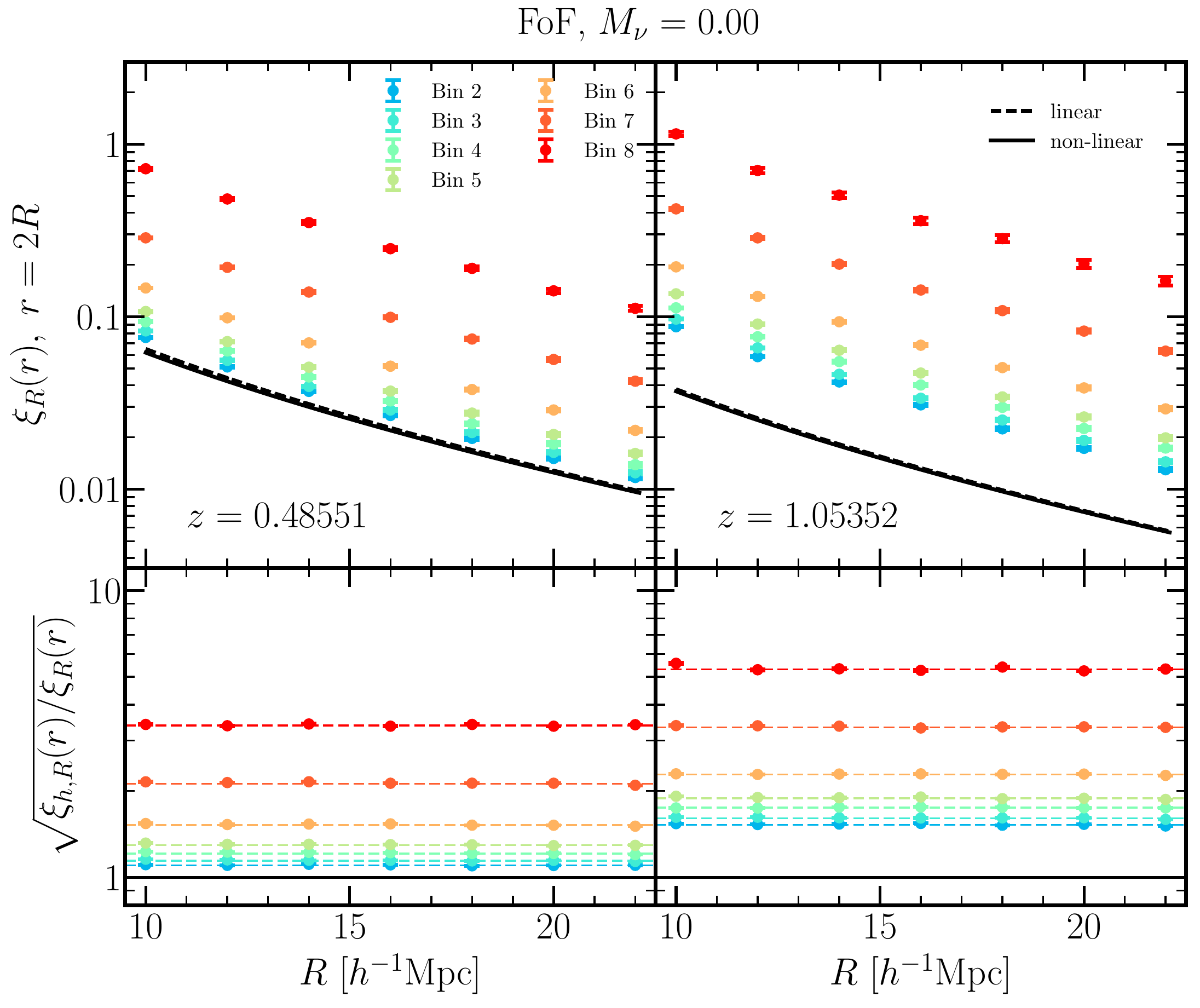}
\includegraphics[width=0.49\textwidth]{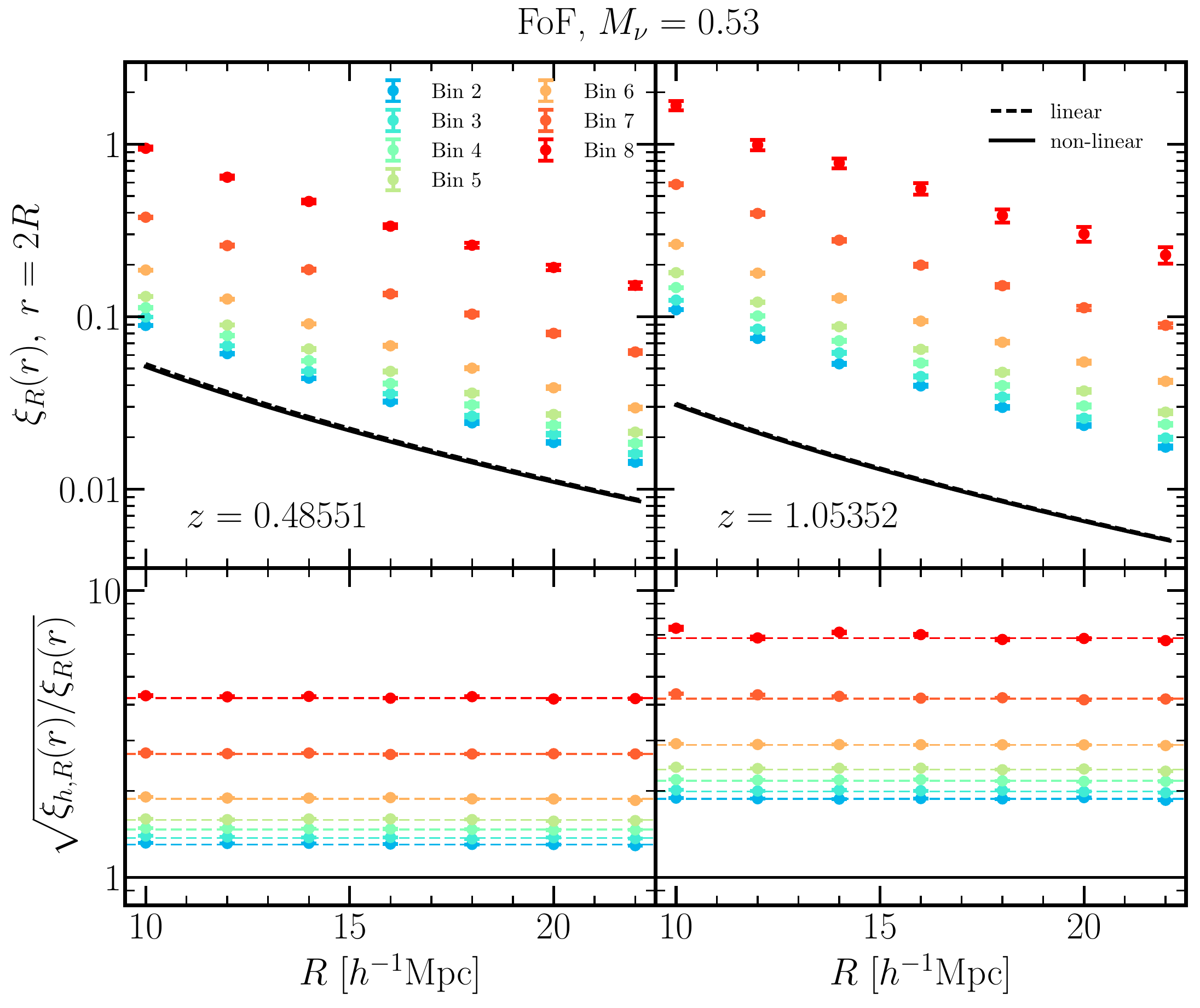}
\caption{Correlation function of the Friend-of-Friends in the simulations with two different cosmologies. In the plot on the left there are the measurements in the $\Lambda$CDM cosmology, while in the plot on the right the $\Lambda$CDM$\nu$ simulation with $M_\nu=0.53$ eV. In both cases, left panel is at redshift $z=0.48551$ and right panel at $z=1.05352$. In the top panel we show the correlation function measured in the mass bins presented in Tab. \ref{tab:mass-bins} (points) compared to the theoretical smoothed matter correlation function (black solid line). In the bottom panel there is the FoF-matter bias, computed as $b=\sqrt{\xi^{FoF}_R(r)/\xi_R(r)}$. The linear bias in the simulation with massive neutrinos is larger than in the standard $\Lambda$CDM case, because, as neutrinos reduce clustering, massive haloes become rarer. We the bias values with a straight line between $R=16$ and $22~h^{-1}$Mpc. As the fit shows, the linear bias is compatible with being constant with scale.}
\label{fig:bias_group}
\end{figure*}

\begin{table}
\centering
\begin{tabular}{| c | c |}
\hline
Bin & Mass range $[10^{12}~h^{-1}~M_\odot]$\\
\hline
0 & $ 0.58 \leq {M} < 1.16 $ \\
1 & $ 1.16 \leq {M} < 2.32 $ \\
2 & $ 2.32 \leq {M} < 3.28 $ \\
3 & $ 3.28 \leq {M} < 4.64 $ \\
4 & $ 4.64 \leq {M} < 6.55 $ \\
5 & $ 6.55 \leq {M} < 9.26 $ \\
6 & $ 9.26 \leq {M} < 30 $ \\
7 & $ 30 \leq {M} < 100 $ \\
8 & $ M \geq 100 $ \\
\hline
\end{tabular}
\caption{Subdivision of the halo catalogues in mass bins.}
\label{tab:mass-bins}
\end{table}
\begin{table*}
\centering
\begin{tabular}{| c | c | c | r | r | r | r | r | r | r | r | r |}
\hline
 & & & bin 0 & bin 1 & bin 2 & bin 3 & bin 4 & bin 5 & bin 6 & bin 7 & bin 8 \\
\hline
\multirow{8}{*}{FoF} & \multirow{4}{*}{$z=0.48551$} & $M_\nu = 0.00$ eV & 0 & 0 & 2902221 & 3509393 & 2402274 & 1708375 & 2878557 & 758008 & 145410 \\ 
  &  & $M_\nu = 0.17$ eV & 0 & 0 & 3152025 & 3178910 & 2430866 & 1667315 & 2712374 & 690356 & 122241 \\ 
  &  & $M_\nu = 0.30$ eV & 0 & 0 & 3116471 & 3269324 & 2263001 & 1642694 & 2589654 & 634844 & 104539 \\ 
  &  & $M_\nu = 0.53$ eV & 0 & 0 & 3273517 & 3026718 & 2144285 & 1501769 & 2334332 & 532432 & 76127 \\
\cline{2-12} 
  & \multirow{4}{*}{$z=1.05352$} & $M_\nu = 0.00$ eV & 0 & 0 & 2571902 & 3039264 & 2003119 & 1358767 & 2044706 & 389064 & 38299 \\ 
  &  & $M_\nu = 0.17$ eV & 0 & 0 & 2713196 & 2674546 & 1958721 & 1277479 & 1836860 & 328973 & 28852 \\ 
  &  & $M_\nu = 0.30$ eV & 0 & 0 & 2620295 & 2674912 & 1767015 & 1218810 & 1679768 & 283298 & 22244 \\ 
  &  & $M_\nu = 0.53$ eV & 0 & 0 & 2619539 & 2335248 & 1570202 & 1036782 & 1386909 & 206588 & 13059 \\ 
\hline
\multirow{8}{*}{SO} & \multirow{4}{*}{$z=0.48551$} & $M_\nu = 0.00$ eV & 453415 & 2973241 & 2783664 & 2361431 & 1669494 & 1210036 & 2064503 & 527389 & 88178 \\ 
&  & $M_\nu = 0.17$ eV & 471602 & 2993986 & 2904400 & 2118437 & 1668026 & 1166042 & 1917029 & 470522 & 72097 \\ 
  &  & $M_\nu = 0.30$ eV & 486007 & 2997527 & 2842491 & 2159522 & 1559049 & 1112575 & 1806821 & 424202 & 60046 \\ 
  &  & $M_\nu = 0.53$ eV & 508713 & 3259563 & 2582711 & 1956593 & 1424500 & 1014066 & 1587591 & 343288 & 41497 \\ 
\cline{2-12}
  & \multirow{4}{*}{$z=1.05352$} & $M_\nu = 0.00$ eV & 363148 & 2449897 & 2433901 & 2030689 & 1373779 & 952365 & 1446087 & 263959 & 22482 \\ 
  &  & $M_\nu = 0.17$ eV & 359806 & 2376103 & 2479355 & 1766898 & 1330746 & 885172 & 1280841 & 218832 & 16462 \\ 
  &  & $M_\nu = 0.30$ eV & 354938 & 2304946 & 2374992 & 1750620 & 1206312 & 819053 & 1157522 & 184575 & 12287 \\ 
  &  & $M_\nu = 0.53$ eV & 341645 & 2359735 & 2069025 & 1500656 & 1039892 & 694588 & 932357 & 130319 & 6834 \\ 
\hline
\end{tabular}
\caption{Population of the 9 mass bins for the Friend-of-Friends (FoF) and spherical overdenities with respect to the critical density (SO) at redshift $z=0.48551$ and $z = 1.05352$.}
\label{tab:pop-bins}
\end{table*}
In Fig. \ref{fig:bias_group} we show the estimated correlation functions of each FoF  sample in the two extreme cases of $M_\nu=0$ eV and $M_\nu=0.53$ eV (the same holds for the SOs as well).

We also represent the corresponding  correlation function of the dark matter field, which is used to estimate the linear bias $b_L$ characterizing each halo sample:
\begin{equation}
b_L \equiv \sqrt{\dfrac{\xi^{FoF}_R(nR)}{\xi_R(nR)}}.
\end{equation} 
We find that our cut in mass does indeed correspond to different tracers, with higher bias for higher-mass objects. However, such different halo populations still show a constant bias with respect to scale, which allows us to fit the measured bias in Fig. \ref{fig:bias_group} with flat lines.

The independence of the FoF-matter bias from scale is confirmed also in the massive neutrino case (Fig. \ref{fig:bias_group}, right). In particular, we note here that the bias is generally higher when considering massive neutrinos. This is due to the fact that, since they smooth the matter distribution, neutrinos make haloes of a given mass rarer then in a standard $\Lambda$CDM cosmology.

\begin{figure*}
\centering
\includegraphics[width=0.49\textwidth]{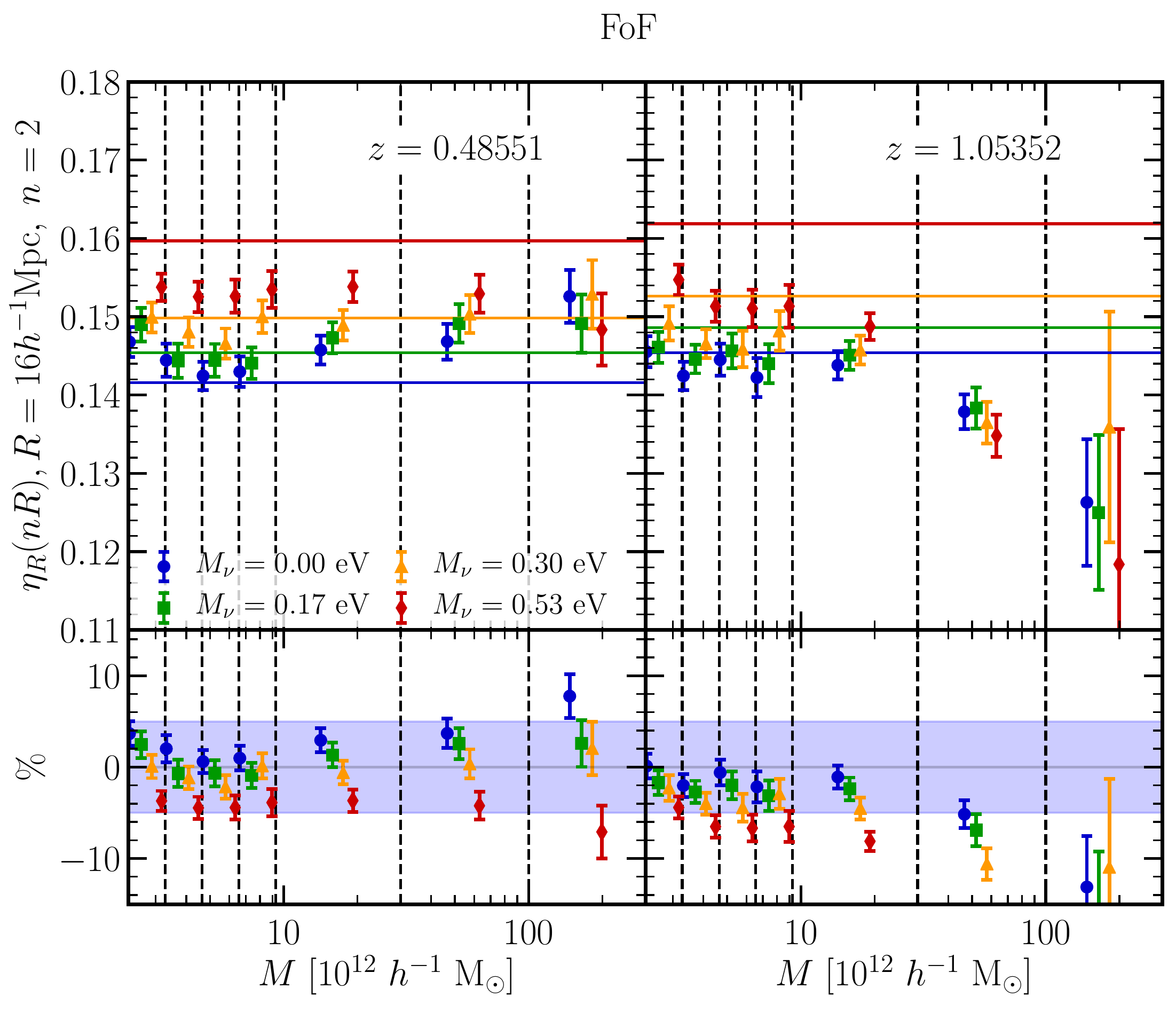}
\includegraphics[width=0.49\textwidth]{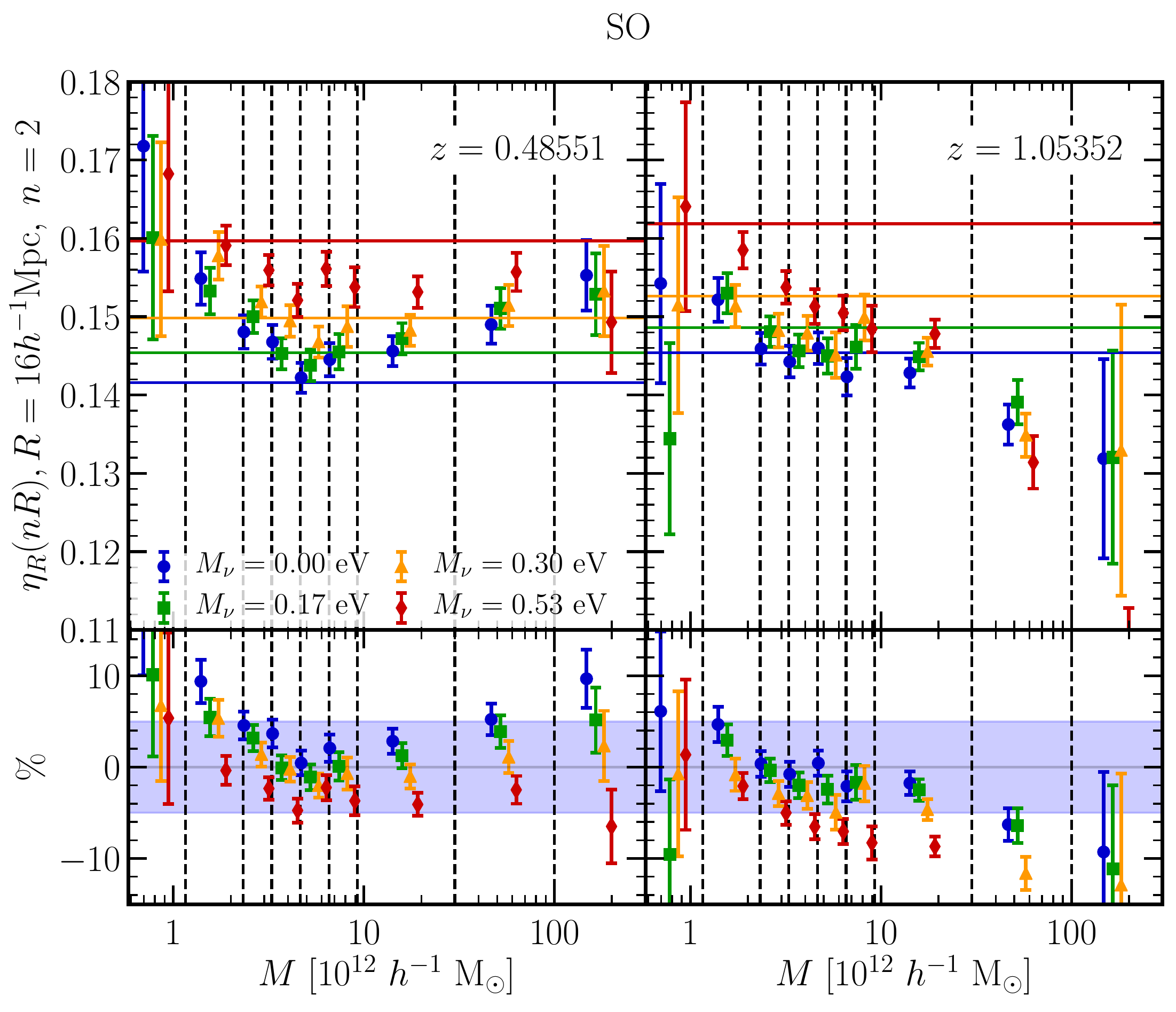}
\caption{The clustering ratio of the Friend-of-Friends (FoF, left plot) and of the spherical overdensities with respect to the critical density (SO, right plot) computed in each mass bin at redshift $z=0.48551$ (left panel of each plot) and $z=1.05352$ (right panel) for all the neutrino masses. The smoothing radius is $R = 16 ~ h^{-1}$ Mpc and the correlation length is twice the smoothing radius. Points are measures in the simulations, while lines are the theoretical predictions. At $z=0.48551$ the agreement between measures and predictions is better than 3\% in the bins with masses $< 30.43 h^{-1} \mathrm{M}_{\odot}$ (i.e. where the bias is $<2$), while in the bins with larger masses, corresponding to larger values of bias, the agreement is at 5\% level. Errors are larger in the highest mass bin because it is more sparsely populated. At $z=1.05352$ the discrepancy between measures and predictions shows a specific dependence on the mass of the tracer. We consider this effect to be due a mass resolution effect in the simulation. Finally we point out that we find similar results for FoF and SO, meaning that the clustering ratio at these scales is insensitive to the linear bias irrespective of the mass tracer we choose.}
\label{fig:mbins-so}
\end{figure*}

Secondly we compute the clustering ratio for both the FoFs and the spherical overdensities at redshift $z=0.48551$ and $z=1.05352$. In each of these cases, we analyse the $\Lambda$CDM simulation, which does not include massive neutrinos, and the $\Lambda$CDM$\nu$ simulations with $M_\nu = \lbrace 0.17,0.30, 0.53 \rbrace$ eV. In the two plots in Fig. \ref{fig:mbins-so} we show the clustering ratio for fixed smoothing radius $R=16~h^{-1}$ Mpc and correlation length $r=2R$ in the same mass bins shown in Tab. \ref{tab:mass-bins} for the FoFs and spherical overdensities respectively. Points are measurements in the simulations, while lines are the predictions obtained from the matter power spectrum. The ratio between measures and prediction is in the bottom panel.
At $z=0.48551$, the measured clustering ratio in the bins with masses $<30.43h^{-1}\mathrm{M}_{\odot}$ agrees with the predictions for the matter at 3\% level. Below $100 ~ h^{-1}\mathrm{M}_{\odot}$ the agreement is within 5\%. For objects with mass greater than this, we observe a more scattered trend. We blame a lower statistical robustness, due to fewer objects falling in these mass bins. In any case, we do not observe any peculiar dependence of the clustering ratio on the mass of the objects, confirming up to a few percent accuracy its independence on the bias at these scales.

Conversely, at $z=1.055$ we do see a dependence of the FoF clustering ratio on the mass. The reason is likely to be a lack of mass resolution in the simulations, as moving towards higher redshifts not enough haloes have formed in the high-mass end of the mass function. For this reason, we expect this situation to be even more sever in the cases with masses neutrinos, which is confirmed in the right power of Fig. \ref{fig:mbins-so}.

As we find similar results for both the FoFs and the spherical overdensities, we conclude that such results do not depend on the tracer we choose to observe.

Finally, we claim that the clustering ratio is insensitive to the bias on linear scales in a cosmology including massive neutrinos irrespective of either the mass of the tracer or the nature of the tracer itself or the total mass of neutrinos considered. This allows us to directly compare real-space predictions of the matter clustering ratio with real-space measures of the clustering ratio of any biased matter tracer, i.e. $\eta_{g,R} (r)\equiv \eta_R(r)$. 

\subsection{Redshift space}
In redshift space, as described in Sec. \ref{sec::cr}, the apparent position of galaxies is modified according to the projection of their peculiar velocity along the line-of-sight. This effect distorts the clustering properties of the distribution and we thus expect its correlation function and variance to be affected. However, we expect redshift-space distortions not to affect the clustering ratio on linear scales (Eq. \ref{eq:CR-identity-zspace}) as the effect is cancelling out into the ratio. In order to verify the accuracy of this approximation we created the redshift space catalogues of the FoFs and spherical overdensities in the simulations, moving the positions of the tracers along an arbitrary direction, chosen as the line-of-sight direction.

Having shown that the clustering ratio does not depend on the way we define the haloes nor on their mass tracer, from now on we focus on the dark matter halo catalogues identified with the Friend-of-Friends algorithm and we compare the two extreme cases of $M_\nu=0$ eV and $M_\nu=0.53$ eV.
Note that we use the $M_\nu=0$ simulation as a reference for comparisons, since it has already been shown that these properties are valid when neutrino are massless \citep[see][]{BelMarinoni2014,BelMarinoniGranett2014}. 

\begin{figure}
\includegraphics[width=0.46\textwidth]{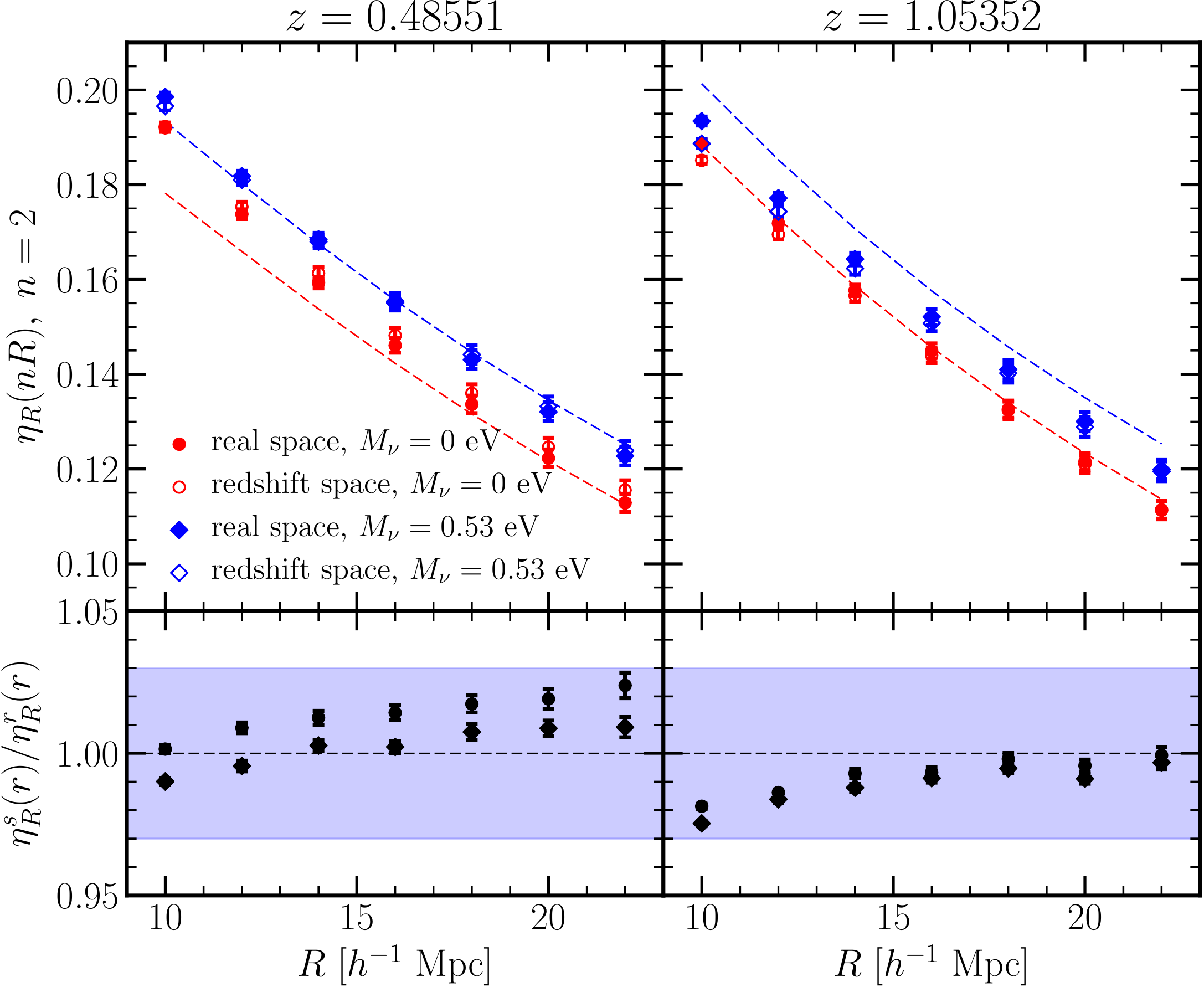}
\caption{The clustering ratio smoothed on the scale $R$ and at correlation length $r=nR,~n=2$ as a function of the smoothing scale. We show in red the measures in the $\Lambda$CDM simulation and in blue the ones in the simulation with the highest neutrino mass, $M_\nu=0.53$ eV, which are the two extreme cases. Filled dots are measures in real space, while empty dots in redshift space. In the bottom panel the ratio between the clustering ratio in redhift space over the real space case is shown. Since on linear scales the monopole contribution coming from redshift-space distortions enhances the correlation function and the variance by the same multiplicative factor, we expect the clustering ratio to be unaffected. The ratio between redshift and real space measurements is, in fact, of order 1 with an accuracy better then 3\%. This is even better confirmed in the case at redshift $z=1.055$ (right) because at higher redshift the matter growth is more linear.}
\label{fig:allmass_group}
\end{figure}

Fig. \ref{fig:allmass_group} shows, at the scales of interest, the independence of the clustering ratio from redshift-space distortions. The ratio between measurements of the clustering ratio in redshift and in real space is of order 1, either with and without massive neutrino, both at $z=0.48551$ and $z=1.05352$. 

In particular we show that we recover the results already obtained in other works \citep{BelMarinoni2012} in the $\Lambda$CDM case. In the case including massive neutrinos we find an agreement between redshift and real space measurements at 3\% level at redshift $z=0.48551$ and at better than 1\% on scales $R \gtrsim 16~h^{-1}$Mpc at $z=1.05352$. In this case the accuracy is higher at higher redshift as the growth of structures is more linear.

Moreover we note that the agreement with predictions is better for the simulation with massive neutrinos with respect to the $\Lambda$CDM one. This is due to the fact that massive neutrinos lower the matter fluctuations (see the values of $\sigma_{8,cc}$ in Tab. \ref{tab::demnuni}) and therefore tend to reduce the velocity dispersion, resulting in redshift-space distortions that are more into the linear regime.

We therefore propose to use the clustering ratio as a cosmological probe to constrain the parameters of the cosmological model. As a matter of fact, the analysis of the simulations has shown that the clustering ratio, besides being independent from the matter tracer and the bias, is not affected by redshift-space distortions on linear scales. This implies that Eq. (\ref{eq:CR-identity-zspace}), $\eta_{g,R}^s(r)\equiv \eta_R$, still holds in the presence of massive neutrinos, allowing us to directly compare clustering ratio measurements in redshift-survey galaxy catalogues to the theoretical matter clustering ratio predictions.
\section{Results}\label{sec::res}

\subsection{Optimisation}
In order to estimate and predict the clustering ratio of galaxies, we need to choose two different scales: the smoothing scale $R$, \textit{i.e.} the radius of the spheres we use for counting objects, and the correlation length $r$, that for simplicity we assume to be some multiple of the smoothing scale, $r=nR$. Choosing the best combinations of $R$ and $r$ is seminal to maximise the information we can extract from this statistical tool.

The smoothing scale $R$ controls the scale under which we make our observable blind to perturbations. A sufficiently large value of $R$ allows us to screen undesired nonlinear effects, that would compromise the effectiveness of the clustering ratio. On the other hand, an excessively large smoothing scale can lead to more noisy measurements, since in the same volume we can accommodate fewer spheres. Moreover, if $R$ is too large, the entire signal would be screened and the measurement would become of little interest.

Also for the correlation length, choosing small values of $R$ and $n$ implies coping with small-scale nonlinearities, which risk to invalidate the identity expressed in Eq. (\ref{eq:CR-identity-zspace}). Large values of correlation distances, however, would make it difficult to accommodate enough couples of spheres in the volume to guarantee statistical robustness. An additional constraint comes from the strategy we adopt to fill the volume with spheres and perform the count-in-cells. In this framework, if the  correlation length is below twice the smoothing scales, $r<2R$, the spheres of our motif of cells would overlap, resulting in an additional shot-noise contribution. For this reason, we only allow values of $n \geq 2$.

\begin{figure}
\centering
\includegraphics[width=0.40\textwidth]{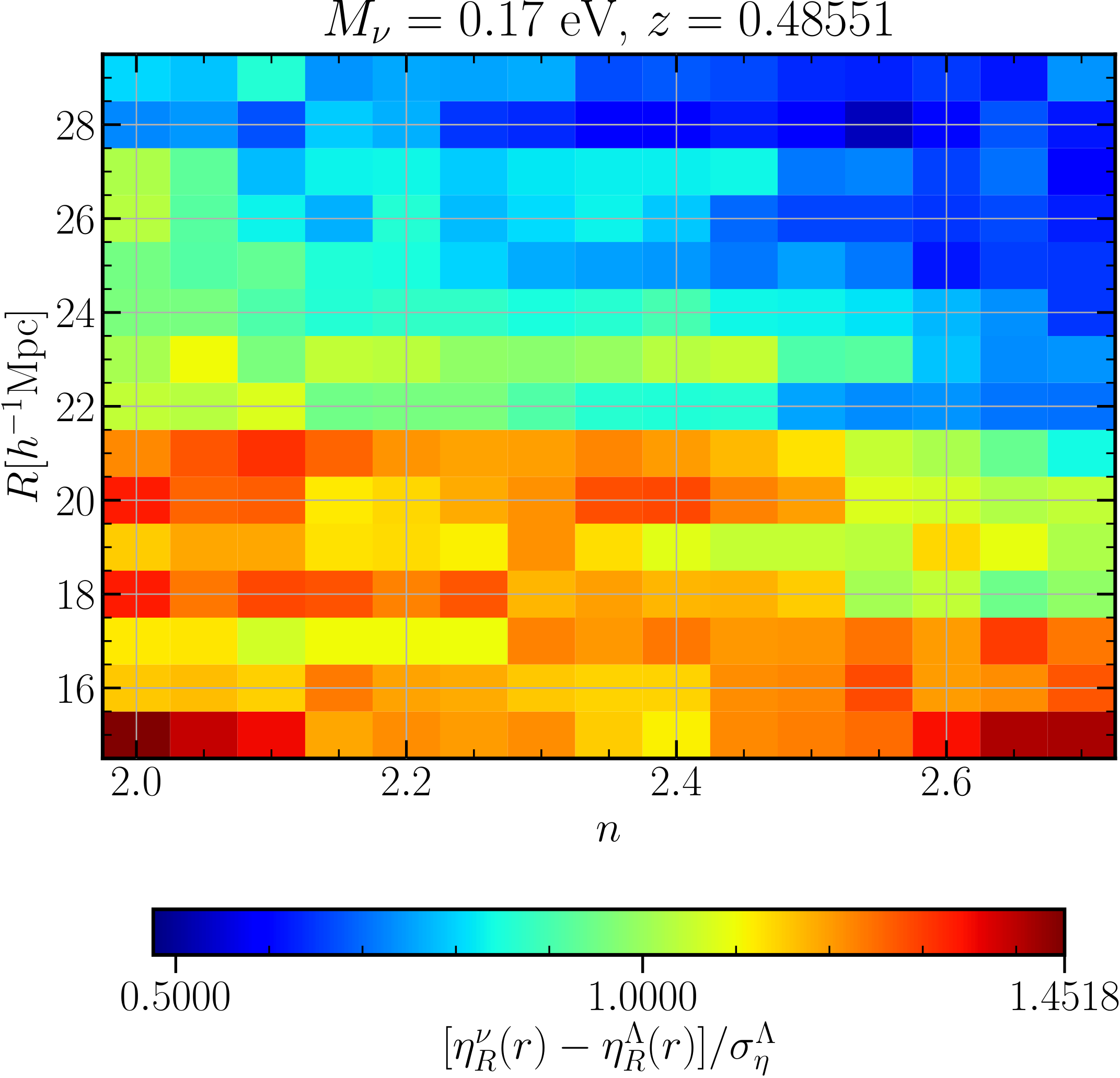}
\caption{The effect of neutrinos on the clustering ratio compared to a $\Lambda$CDM cosmology. In the $(n,R)$ plane, we plot colour contours corresponding to $(\eta_R(r,\nu) - \eta_R(r,\Lambda \mathrm{CDM})/\sigma_{\eta}(\Lambda \mathrm{CDM)}$. As expected, the sensitivity to the neutrino total mass increases at small smoothing scales and correlation lengths (red regions).}
\label{fig::s/n}
\end{figure}

The main information we want to extract is the total neutrino mass. The sensitivity of the clustering ratio to this parameter can be quantified as an effective signal-to-noise ratio, defined as
\begin{equation}
\mathrm{S/N} = \dfrac{\eta_R^\nu(r)-\eta_R^\Lambda (r)}{\sigma_R^\Lambda},
\label{eq::s/n}
\end{equation}
where $\eta_R^\nu(r)$ is the clustering ratio measured in a simulation with neutrino mass $M_\nu$, $\eta_{R}^\Lambda(r)$ is measured in the reference \lcdm{} simulation and $\sigma_{R}^\Lambda$ is the uncertainty on the clustering ratio measured in the \lcdm{} simulation. This quantity measures how much a massive neutrino cosmology is distinguishable from a \lcdm{} one, given the typical errors on the measures of the clustering ratio for the specific volume and number density of tracers, as a function of $R$ and $r$. Fig. \ref{fig::s/n} shows the $(n,R)$ plane, constructed as a grid with correlation lengths $n \in [2,2.75]$  with step $\Delta n = 0.05$ and smoothing scales $R \in [15,30]$ with step $\Delta R = 1$, all distances being expressed in units of $h^{-1} ~ {\rm Mpc}$. At each point on the grid a color is associated, representing the value of this effective signal-to-noise ratio. The effect of massive neutrinos is, as expected, appreciable on small scales (both small $n$ and small $R$), and eventually becomes negligible moving towards large scales.

\begin{figure}
\centering
\includegraphics[width=0.40\textwidth]{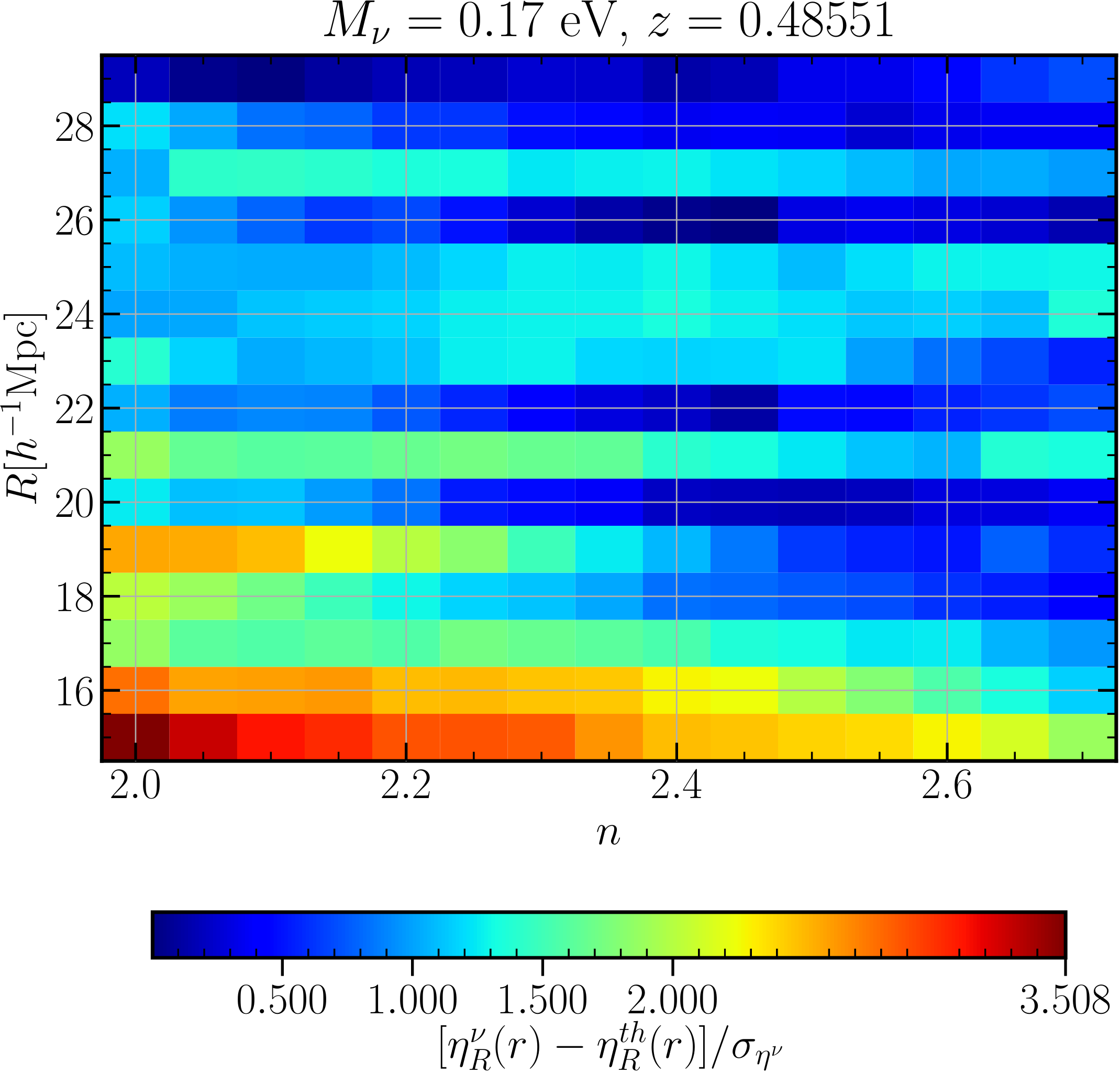}
\caption{Discrepancy between the clustering ratio measured in the simulation and the theoretical prediction in the $(n,R)$ plane. Colours represent the quantity $(\eta_R(r) - \eta_{R}^{\rm th}(r)) / \sigma_{R}^(r)$, the blue regions being the ones with the best agreement with the predictions. Sufficiently large smoothing scales screen the effects of the nonlinear growth of perturbations, allowing us to exploit the clustering ratio as a cosmological probe. By smoothing our distribution on scales $R>19~h^{-1}$ Mpc we ensure an agreement with the model better then $\sim 1.5$ standard deviations.}
\label{fig::th_err}
\end{figure}

While we want to maximise the effects of neutrinos, we want to minimise errors. In particular, we can define a theoretical error, that accounts for the combinations of smoothing radii $R$ and correlation lengths $r$ where the assumptions under which we can apply the identity expressed in Eq. (\ref{eq:CR-identity-zspace}) break down. Such theoretical error can be quantified as 
\begin{equation}
\delta_{\rm th} = \dfrac{\eta_R(r) - \eta_{R}^{\rm th}(r)}{\sigma_{\eta}},
\end{equation}
where $\eta_R(r)$ is the clustering ratio measured in the simulation with a given cosmology, $\eta_R^{\rm th}(r)$ is the prediction obtained with a Boltzmann code, and $\sigma_\eta$ the uncertainty on the measurement. In Fig. \ref{fig::th_err} we show as a color map the values that we obtain for this theoretical error in the same $(n,R)$ plan introduced above. We can see that on very small scales the effect of nonlinearities is not negligible, and we cannot use the clustering ratio as an unbiased observable.

From Fig.s \ref{fig::s/n} and \ref{fig::th_err} we see that we need to balance between the requirement coming from the signal-to-noise ratio (that is maximum on small scales), and those from the theoretical errors (that is minimum on large scales). We introduce, therefore, a way of combining these pieces of information into a single colour map, which we use to seek the sweet-spots, in this parameter space, where both conditions are satisfied. 

\begin{figure}
\centering
\includegraphics[width=0.40\textwidth]{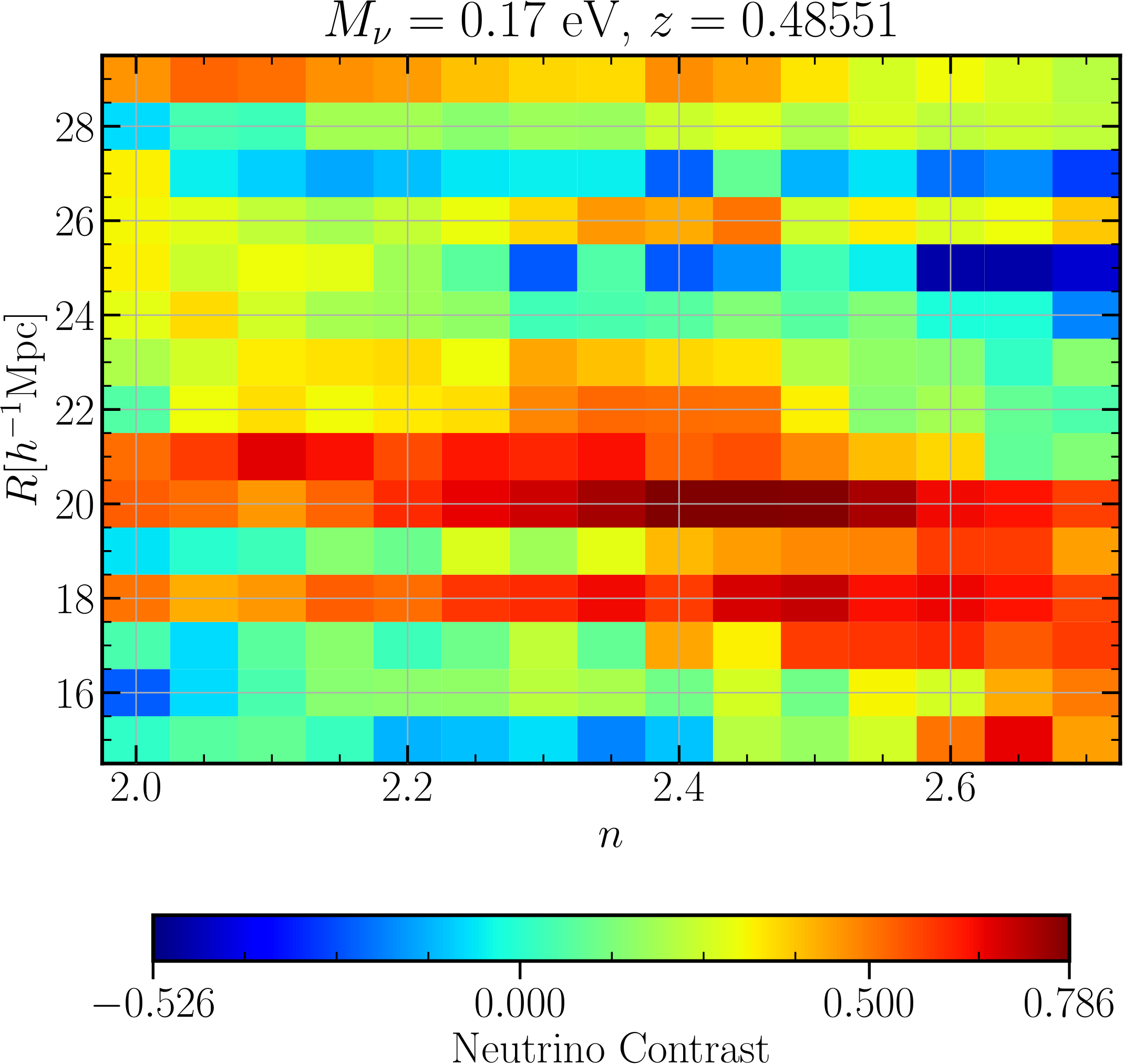}
\caption{In this colour plot we subtract to the neutrino signal-to-noise normalised to 1 a combined error normalised in the same fashion. Details on the definition are in the text. We are interested in the regions in the $(n,R)$ plane where the neutrino signal-to-noise dominates ($\sim 1$) on the error ($\sim 0$), graphically visible as hot spots. The region with smoothing scales $20 < R < 23 ~ h^{-1} \, \mathrm{Mpc}$} seems to be the most promising. In particular, by repeating this test for different redshifts and neutrino masses, we chose as our candidate scales $R=22~h^{-1} \, \mathrm{Mpc}, \, n=2.1$.
\label{fig::contrast}
\end{figure}

First, we define a combined percentage error (that accounts both for statistical errors and discrepancies from the model) as
\begin{equation}
\delta_{\rm combined} = \left\lbrace \dfrac{\eta_R^\nu(r) - \eta_R^{\nu,th}(r)}{\eta_R^\nu(r) - \eta_R^\Lambda(r)} \right\rbrace \left\lbrace \dfrac{\sigma_\eta^\Lambda}{\eta_R^\nu(r) - \eta_R^\Lambda(r)} \right\rbrace.
\label{eq::combinederr}
\end{equation}
The quantity in the first parenthesis is related to how much the statistical error is important with respect to the effects of neutrinos, while the second parenthesis is a weight that accounts for the typical uncertainty on the measurement in each bin of $n$ and $R$. Finally, we define the neutrino contrast as
\begin{equation}
C = \dfrac{\mathrm{S/N}}{\mathrm{max(S/N)}} - \dfrac{\delta_{\rm combined}}{\mathrm{max}(\delta_{\rm combined})}.
\end{equation}
Here we have normalised the signal/noise defined in Eq. (\ref{eq::s/n}) and the combined error defined in Eq. (\ref{eq::combinederr}) to their respective maxima (on the considered grid) and we are interested in finding the regions where this contrast is dominated by the signal/noise, \textit{i.e.} where $C(n,R) \sim 1$. In Fig. \ref{fig::contrast} we show the neutrino contrast on the $(n,R)$ grid, for the simulation with $M_\nu=0.17$ eV at redshift $z=0.48551$.

We have repeated this analysis for the three massive neutrino simulations (with $M_\nu=0.17,\,0.30,\,0.53$ eV) and for different redshifts, spanning the range from $z=0.48551$ to $z=2.05053$. Our conclusion is that the combination $R=22~ h^{-1} \, \mathrm{Mpc}$, $n=2.1$ is the most viable candidate for all these cosmologies and redshifts.

\subsection{Likelihood}

We aim at comparing measurements and predictions of the clustering ratio, in order to find the set of parameters of the model that maximizes the likelihood. We exploit the different dependence of measurements and predictions on the cosmological model. In particular, the measured value of the clustering ratio depends on the way we convert redshifts and angles into comoving distances, that depends on the total matter density $\Omega_m$, on the dark energy fraction $\Omega_\Lambda$ and on the background expansion rate $H(z)$.

On the other hand, the theoretical prediction for the clustering ratio depends on the entire cosmological model and is therefore sensitive also to the value of the total neutrino mass $M_\nu$.

We choose six baseline free parameters in our analysis, namely the baryon and cold dark matter density parameters $\Omega_b h^2$ and $\Omega_{cdm}h^2$, the Hubble parameter $H_0$, the optical depth at the recombination epoch $\tau$, the amplitude of the scalar power spectrum at the pivotal scale $A_s$ and the scalar spectral index $n_s$. Moreover, we extend this parametrization with two additional free parameters, the total neutrino mass $M_\nu$ and the equation of state of the dark energy fluid $w$. The most general vector of parameters therefore is
\begin{displaymath}
\boldsymbol{p} = \lbrace \Omega_{b}h^2, \Omega_{cdm}h^2, H_0, \tau, A_s, n_s, M_\nu, w \rbrace.
\end{displaymath}
We follow \cite{BelMarinoni2014}, who showed that the likelihood function of the clustering ratio (given a fixed set of parameters) is compatible with being a Gaussian. Therefore we will compute the logarithmic likelihood as $\ln \mathcal{L} = - \chi^2 / 2$ (apart from a normalization term) where 
\begin{equation}
\chi^2 (\boldsymbol{p}) = \sum_i \dfrac{ (\eta_{R,i} (r) - \eta_{R,i}^{\mathrm{th}} (r))^2 }{ \sigma_{\eta_i}^2 }, 
\end{equation}
where we neglect the covariance between the different redshift bins.

We account for the dependence of the measurements on the cosmological model assumed, induced by the cosmology-dependant conversion of redshifts into distances, whenever we compare measurements of the clustering ratio (obtained in the fiducial cosmology) to its predictions (in a generic cosmology). That is, when computing the likelihood for the set of parameters $\boldsymbol{\vartheta}$, we must keep in mind that the measured value has been computed in a different cosmology, the one with the fiducial set of parameters $\boldsymbol{\vartheta}^{\mathrm{F}}$.

We keep the measurements fixed in the fiducial cosmology and rescale the predictions accordingly. We consider that, due to Alcock-Paczy\'nski effect, at same redshift and angular apertures we can associate different lengths depending on cosmology \citep{AlcockPaczynski1979}.

Our measurements depend on distances only through the smoothing scale $R$. This is because the correlation length is always expressed as a multiple of the smoothing scale, $r=nR$. This means that, since the measure has been obtained in the fiducial cosmology using spheres of radius $R^{\mathrm{F}}=22~h^{-1}$Mpc, they need to be compared to predictions obtained in a generic cosmology using a smoothing length $R=\alpha R^{\mathrm{F}}$, where $\alpha$ is our Alcock-Paczy\'nski correction.

We write the Alcock-Paczy\'nski correction $\alpha$ as \citep{EisensteinEtal2005}

\begin{equation}
\alpha = \left[ \dfrac{E^\mathrm{F}(z)}{E (z)} \left( \dfrac{D_{\mathrm A}}{D_{\mathrm A}^\mathrm{F}} \right)^2 \right]^{1/3},
\end{equation} 
where $E(z) \equiv H(z)/H_0$ is the normalised Hubble function and $D_{\rm A}$ the angular diameter distance. Therefore, we are going to compare
\begin{equation}
\eta_{g,R}^{F,s}(nR) \equiv \eta_{\alpha R} (n \alpha R),
\label{eq::identity-ap}
\end{equation}
the left hand side of Eq. (\ref{eq::identity-ap}) being the clustering ratio of galaxies measured in redshift space assuming the fiducial cosmology, while the right hand side is the predicted clustering ratio for matter in real space, rescaled to the fiducial cosmology to make it comparable with observations.

In order to efficiently explore the parameter space we have modified the public code \texttt{CosmoMC} \citep{LewisBridle2002}, adding a likelihood function that implements this procedure.

\subsection{Constraints using SDSS data}

We measure the clustering ratio in the 7$^{\rm th}$ \citep{AbazajianEtal2009} and 12$^{\rm th}$ \citep{AlamEtal2015} data release of the Sloan Digital Sky Survey (SDSS) by smoothing the galaxy distribution with spherical cells of radius $R$ and counting the objects falling in each cell. We divide the sample into three redshift bins that have mean redshifts $\bar{z}=\lbrace 0.29,0.42,0.60 \rbrace$. The first redshift bin is extracted from the DR7 catalogue, while the two bins at higher redshift come from the DR12 catalogue, after removing the objects already present in the other bin. 

To perform the count-in-cell procedure, we convert redshifts into distances, assuming a cosmology with $H_0 = 67 \mathrm{\,km\, s^{-1}\, Mpc^{-1}}$, $\Omega_m = 0.32$ and in which we fix the geometry of the universe to be flat, $\Omega_k = 0$, forcing $\Omega_\Lambda=1- \Omega_r - \Omega_m$. Therefore, this is to be considered our fiducial cosmology. We compute the clustering ratio using the estimators presented in Sec. \ref{subsec:estimators}, employing our optimised smoothing size and correlation length, $R=22 ~ h^{-1}$Mpc and $r = 2.1~R$. Our measures of the clustering ratio in each redshift bin are
\begin{equation}
\begin{split}
{\rm at~0.15\leq z \leq 0.43},\quad &\eta_{g,R}(r) = 0.0945 \pm 0.0067, \\
{\rm at~0.30\leq z \leq 0.53},\quad &\eta_{g,R}(r) = 0.0914 \pm 0.0055, \\
{\rm at~0.53\leq z \leq 0.67},\quad &\eta_{g,R}(r) = 0.1070 \pm 0.0110.
\end{split}
\nonumber
\end{equation}
Details on the computation of the clustering ratio and its errors in the SDSS catalogue can be found in \cite{BelHoffmannGaztanaga2015}, where, though, measures are performed assuming a different fiducial cosmology.

\begin{figure}
\includegraphics[width=0.48\textwidth]{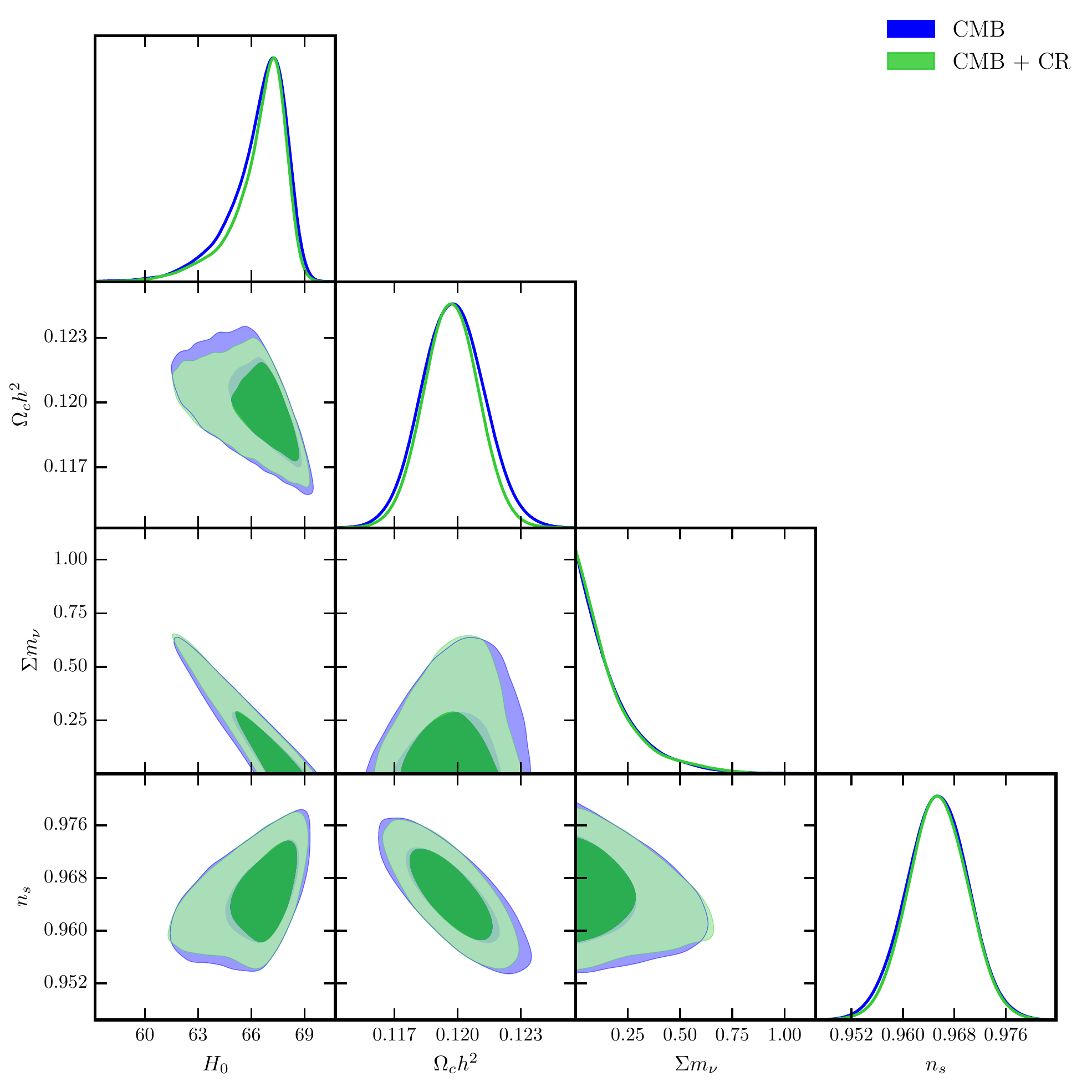}
\caption{Joint posterior distribution obtained using Planck temperature and polarization data and the clustering ratio measured in SDSS DR7 and 12. We fit a cosmological model with seven free parameters, the six baseline parameters of Planck and the neutrino total mass, but here only four of them are shown.}
\label{fig::sdss_data_tri}
\end{figure}

In Fig. \ref{fig::sdss_data_tri} we show, for some relevant parameters, the joint posterior distribution obtained fitting at the same time the Planck temperature and polarization data and the clustering ratio measurements in SDSS DR7 and DR12, leaving free to vary the six baseline parameters and the total neutrino mass $M_\nu$. Already by eye, adding the clustering ratio to the CMB information does not seem to improve much the upper bound of the total neutrino mass parameter. In general, the most significant improvement seems to occur on the constraint of the cold dark matter density parameter.

\begin{figure*}
\includegraphics[width=\textwidth]{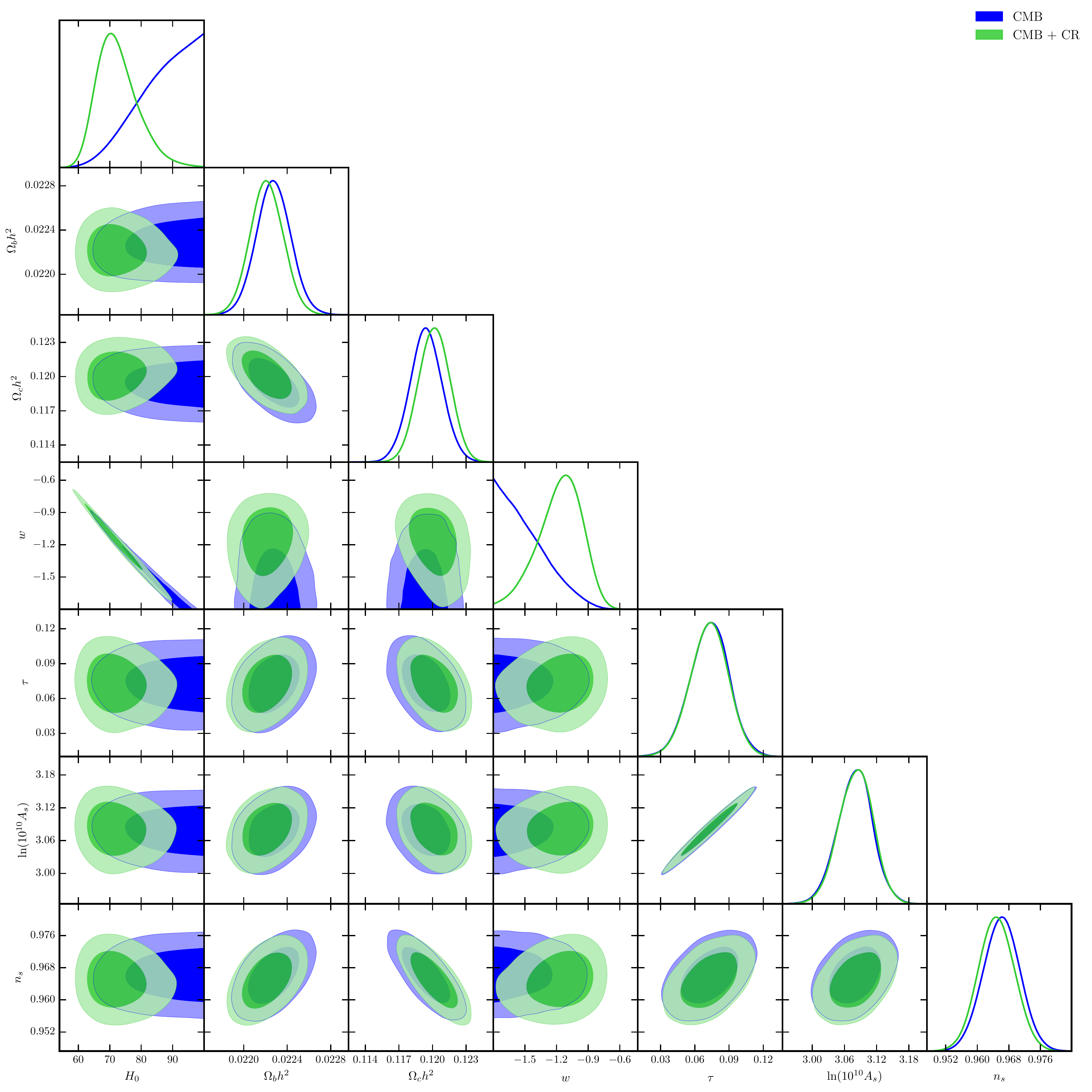}
\caption{Joint posterior distribution obtained employing CMB temperature and polarization data from Planck and the clustering ratio measurements from SDSS DR7 and DR12 catalogues. Besides the six standard parameters of the model, also the equation of state of dark energy is left free.}
\label{fig::w}
\end{figure*}

Moreover, we have also checked how constraints change when we leave the equation of state of dark energy, $w$, as an additional free parameter. As a matter of fact, $w$ is known to be strongly degenerate with the other parameters of the model, when only CMB data are used. In general, we need information from a geometrical probe sensitive to the late time universe in order not to find non-physical solutions. Fig. \ref{fig::w} shows that the clustering ratio is indeed able to break such degeneracy.

Also the combination of other cosmological probes can help breaking degeneracies and tightening constraints. For this reason we compare the constraining power of the clustering ratio to that of two other observables, the fit of the BAO peak in the correlation function measured by the BOSS collaboration in the DR11 CMASS and LOWZ datasets \citep{AndersonEtal2014} and the lensing of the CMB signal due to the intervening matter distribution between the last scattering surface and us,  where the amplitude of the lensing potential, $A_L$, has been kept fixed to 1 \citep{PlanckLikelihoods2016}.

\begin{figure*}
\centering
\includegraphics[width=0.40\textwidth]{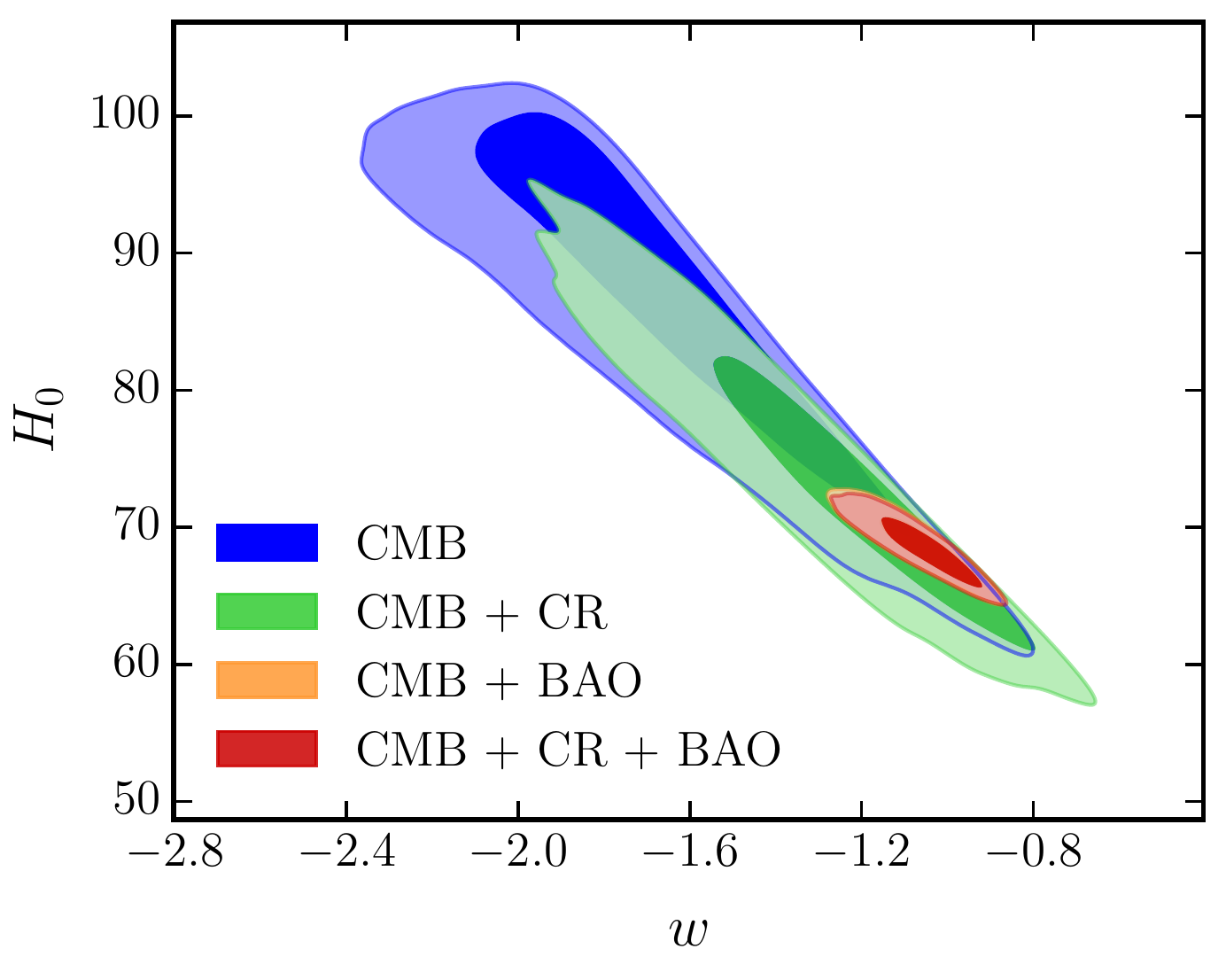}
\includegraphics[width=0.40\textwidth]{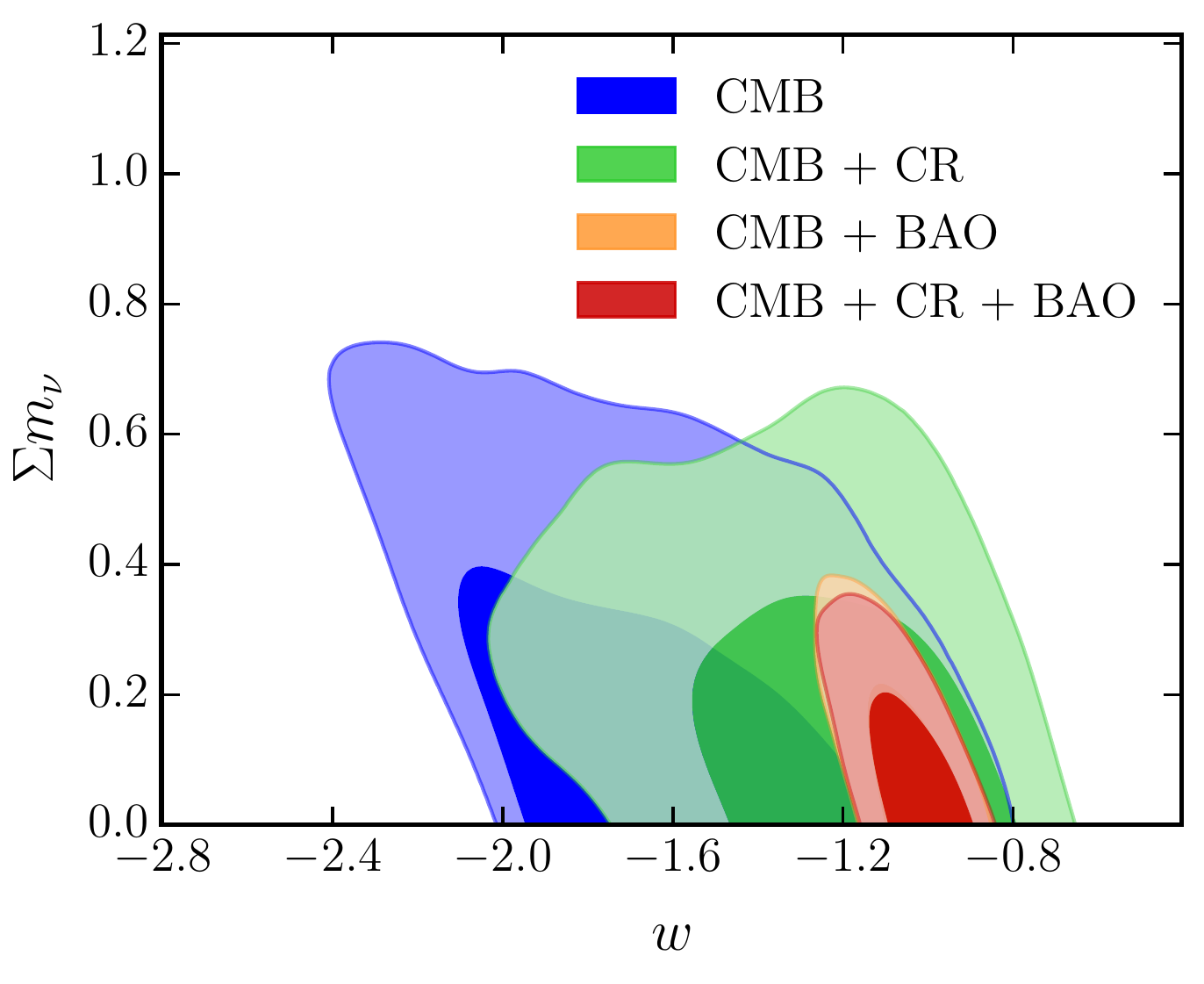}
\caption{Degeneracy between the equation of state of dark energy, $w$, and, on the left, the Hubble parameter today $H_0$, and on the right the total neutrino mass $M_\nu$. The considered likelihoods are the one computed using Planck data alone, and its combinations with the clustering ratio measured in SDSS DR7 and 12, the BAO position from SDSS DR11 or both.}
\label{fig::2d_w_H_bao}
\end{figure*}

\begin{figure*}
\centering
\includegraphics[width=0.40\textwidth]{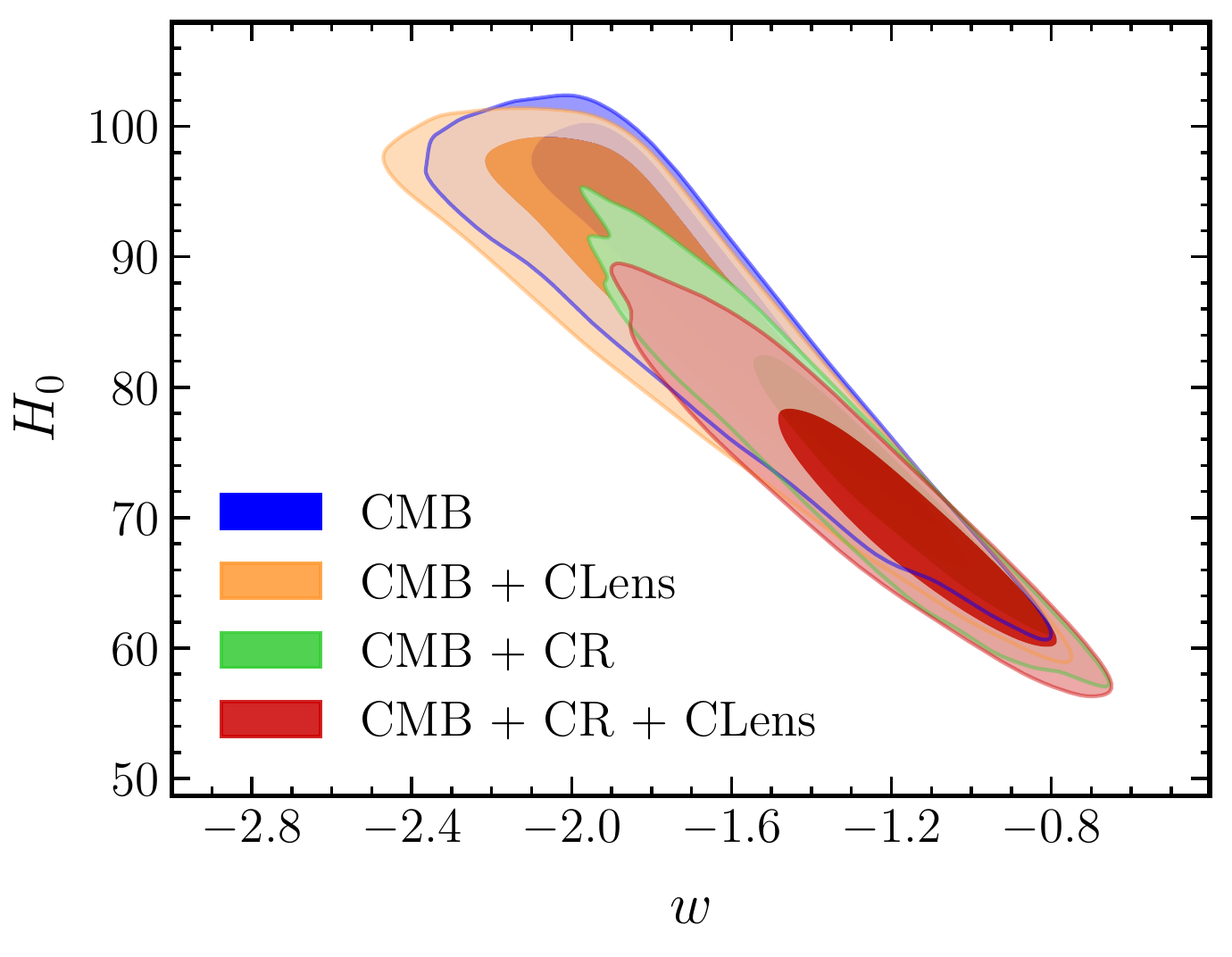}
\includegraphics[width=0.40\textwidth]{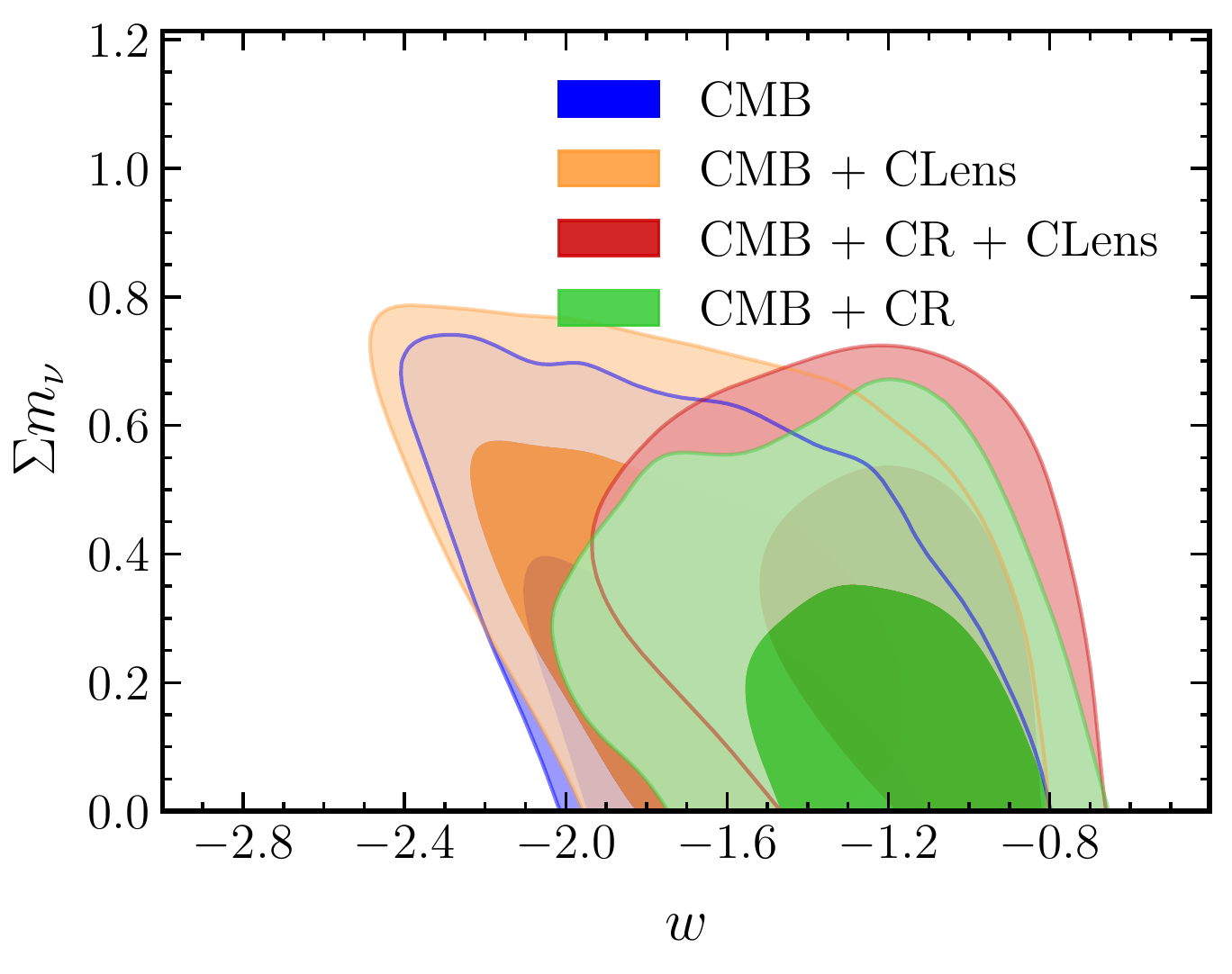}
\caption{Degeneracy between the equation of state of dark energy, $w$, and, on the left, the Hubble parameter today $H_0$, and on the right the total neutrino mass $M_\nu$. The considered likelihoods are the one computed using Planck data alone, and its combinations with the clustering ratio measured in SDSS DR7 and 12, the lensing of the CMB or both.}
\label{fig::2d_w_mnu_clens}
\end{figure*}

In the first part of Tab. \ref{tab::like_results} we show the mean, 68\% and 95\% levels obtained for the different parameters combining the likelihoods presented above. To better show the behaviour of the clustering ratio with respect to the other probes considered, in Fig.s \ref{fig::2d_w_H_bao}-\ref{fig::2d_w_mnu_clens}, we focus particularly on the parameters $w$, $M_\nu$ and $H_0$.

In general, adding the clustering ratio considerably improves on the parameter constraints obtained with CMB data alone, especially when also $w$ is free to vary. In particular, the clustering ratio is able to break the degeneracy between $w$ and the other cosmological parameters, that affects the constraints drawn with the sole CMB data. On the other hand, the clustering ratio does not seem to improve much the constraint on the $M_\nu$ parameter, especially when compared to probes such as the CMB lensing and the BAO peak position.

The clustering ratio proves to be extremely sensitive to the cold dark matter fraction $\Omega_{cdm}h^2$, as adding the clustering ratio to the CMB analysis results in a 12\% improvement on the 95\% confidence level.

To improve our understanding of the results presented in the previous section, we investigate how well the clustering ratio allows us to recover a certain known cosmology. 

To this purpose, we use the measurements of the clustering ratio in one of the DEMNUni simulations, the one with $M_\nu = 0.17$ eV, which represents the closest value to the current available constraints on the neutrino total mass. The clustering ratio is measured in the simulation at the same redshifts, and with the same binning, as in the SDSS data. The error on each measurement in the simulation is taken to be the one obtained from the SDSS measurements.

The likelihood using the CMB data is computed in this case fixing the bestfits to the values of the parameters in the cosmology of the simulation, and employing the covariance matrix contained in the publicly available Planck data release.

\begin{figure}
\includegraphics[width=0.48\textwidth]{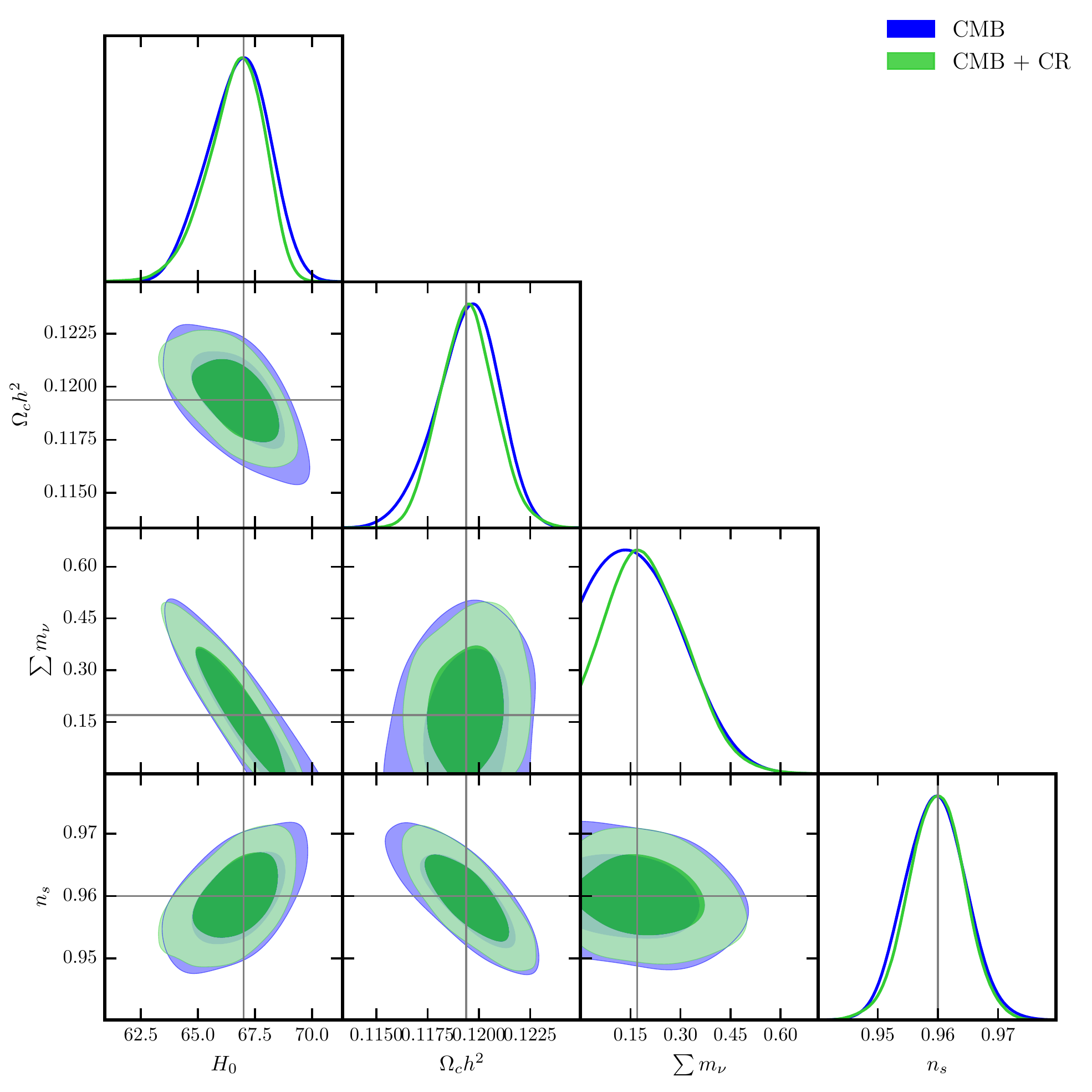}
\caption{Joint posterior distribution obtained using Planck temperature and polarization data and the clustering ratio. Bestfits here are fixed, errors for Planck come from the publicly available covariance matrix, errors on the clustering ratio have been compute in SDSS DR7 and DR12. Four of the seven free parameters are shown.}
\label{fig::sdss_sim_tri}
\end{figure}

The posterior distribution obtained with this procedure is shown in Fig. \ref{fig::sdss_sim_tri}, while the second part of Tab. \ref{tab::like_results} summarizes the improvements on the constraints on the parameters that we obtain adding the clustering ratio.
We correctly recover the bestfits of our known cosmology, with errors comparable with the true ones. We conclude that the reason why we did not achieve a significant improvement on the constraint on the total neutrino mass using the SDSS data resides in the fact that the clustering ratio requires smaller error bars to be effective in constraining such parameter. We can therefore expect that, with upcoming, large galaxy redshift surveys, the clustering ratio will reach a larger constraining power.

We test such an hypothesis in the next section, analysing the clustering ratio expected for a Euclid-like galaxy redshift survey, in combination with CMB data.

\subsection{Forecasts for a Euclid-like galaxy redshift survey}
In order to forecast the constraining power of the clustering ratio, expected from a future, Euclid-like galaxy redshift survey, we construct the synthetic clustering ratio data in the following way:

\begin{itemize}
\item We imagine to have 14 redshift bins, from $z=0.7$ to $z=2$, with $\Delta z=0.1$
\item In each redshift bin, the synthetic measurement of the clustering ratio is given by the predicted clustering ratio (computed using a Boltzmann code), to which we add a small random noise (within 1 standard deviation).
\item We measure the errors (at the same redshifts) in the DEMNUni simulations; the errors in the simulations are then rescaled, according to the operative formula presented below, to match the volume and number density of our Euclid-like survey.
\end{itemize}

The relative error on the clustering ratio depends on the volume and number density of the sample, and can be parametrised, following \cite{BelHoffmannGaztanaga2015}, as
\begin{equation}
\dfrac{\delta \eta}{\eta} = A V^{-1/2} \exp \left\lbrace 0.14 \left[ \ln \rho - \dfrac{\ln^2 \rho}{2\ln (0.02)} \right] \right\rbrace
\end{equation}
where $V$ is the volume expressed in $h^{-3}\mathrm{Mpc}^3$, $\rho$ is the object number density in $h^3 \mathrm{Mpc}^{-3}$ and $A$ is a normalization factor computed with the reference volume and number density.

\begin{figure*}
\includegraphics[width=\textwidth]{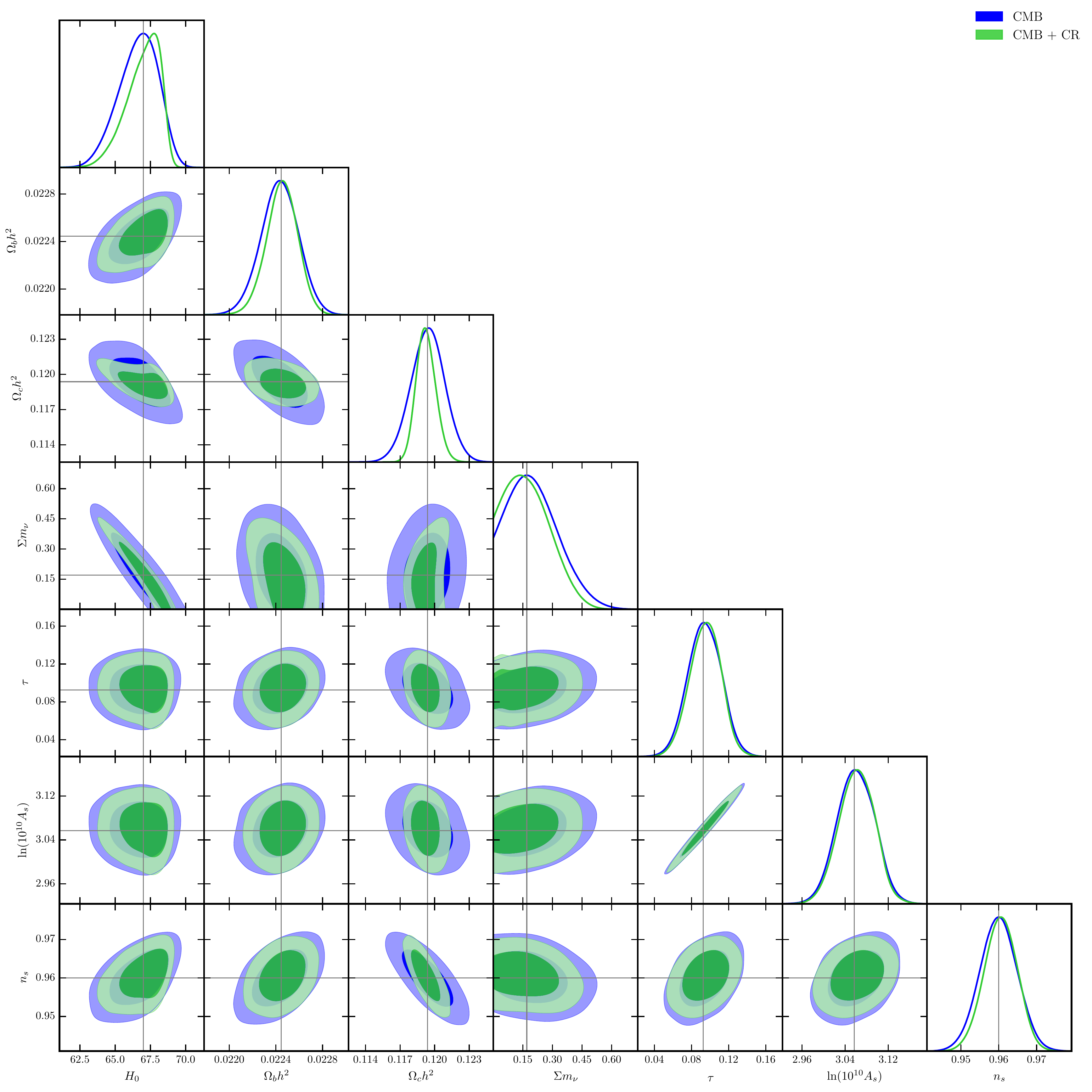}
\caption{Joint posterior distribution obtained combining Planck temperature and polarization data and the clustering ratio measured in a Euclid-like galaxy survey. For CMB data, errors come from the publicly available covariance matrix, for the clustering ratio errors have been measured in the DEMNUni simulations and rescaled to match the volume and number density of our mock survey.}
\label{fig::euclid_tri}
\end{figure*}

We use these data to explore the posterior distribution of the parameters of the model. The results are shown in Fig. \ref{fig::euclid_tri}, and the constraints are shown in the last part of Tab. \ref{tab::like_results}.

In this case there is a much larger improvement on the constraints of all the parameters. The neutrino total mass parameter improves by 14\% on the 95\% limit with respect to using Planck alone. Most notably, the constraint on the cold dark matter density parameter, $\Omega_{cdm}h^2$, improves by over 40\%. Also the spectral index $n_s$ shows a 10\% improvement and the constraint on the Hubble constant $H_0$ improves by 20\%.

This means that, when new data, covering a larger volume, will be available, clustering ratio measurements are expected to contribute with a significant improvement on the constraints on the parameters of the cosmological model.

We also note that, as more different observations are carried out, it becomes very interesting to enhance the constraining power of the clustering ratio also combining its measurements in different datasets. This can be easily done since the clustering ratio is a single measurement, thus scarcely dependent on the survey geometry.
\section{Summary and Conclusion}\label{sec:conclusion}
Neutrino effects are being increasingly included in cosmological investigations, becoming in fact part of the standard cosmological model. Thanks to these investigations, the description of the statistical properties of the universe is gaining the precision required by forthcoming experiments and, at the same time, neutrino physics gains tighter constraints.

In this work we have considered the clustering ratio, an observable defined as the ratio between the smoothed correlation function and variance of a distribution, and extended its range of applicability to cosmologies that include a massive neutrino component. As a matter of fact, the clustering ratio, which has already been tested in $\Lambda$CDM cosmologies including only massless neutrinos, is unbiased and independent from redshift-space distortions on linear scales. As massive neutrinos introduce characteristic scale dependencies in the clustering of galaxies (and matter), such peculiar properties of the clustering ratio needed to be confirmed (or denied) in this cosmological framework.

We divided our analysis into two steps: first, we studied the properties of the clustering ratio in simulations that include massive neutrinos; afterwards, we used the clustering ratio to compute the likelihood of the parameters of the cosmological model, using both real data and forecasts of future data.

In the first part of this work, we employed the DEMNUni simulations to test the clustering ratio in the presence of massive neutrinos. These are the largest available simulations that include massive neutrinos as a separate particle species along with cold dark matter. We computed the clustering ratio using different tracers (dark matter FoF haloes and spherical overdensities), divided into different mass bins (spanning  the interval from $\sim 6 \times 10^{11}$ to $\gtrsim 10^{14}~h^{-1}{\rm M}_{\odot}$), and we explore different choices of  neutrino mass ($M_\nu=\{0,0.17,0.3,0.53\}$ eV) in real and redshift space.

From such analysis we conclude that the properties of the clustering ratio hold also in cosmologies with massive neutrinos. In particular its main property, the fact that the galaxy clustering ratio in redshift space is directly comparable to the clustering ratio predicted for matter in real space on a range of linear scales, is proven valid.

We have therefore moved to employing the clustering ratio as a cosmological probe to find the set of parameters of the model that maximizes the likelihood function, given a set of data. We have used the data from the SDSS DR7+DR12 catalogue. We have computed the clustering ratio in three redshift bins and used these measures in combination with the temperature and polarization anisotropies of the CMB measured by the Planck satellite to explore the likelihood in parameter space with an MCMC approach. We find that the clustering ratio is able to break the degeneracy, present in the CMB data alone, between the equation of state of dark matter, $w$, and the other parameters. Moreover it improves the 95\% limit on the CDM density parameter by $\sim 12\%$. However, we do not find an appreciable improvement in the constraint on the neutrino total mass.

By analysing simulations we conclude that we blame such lack of improvement on the statistical errors, which, with current data, are not yet competitive enough. We have therefore tested the constraining power of the clustering ratio using the error bars expected from a Euclid-like galaxy survey.

In this case we find that not only does the clustering ratio greatly improve (with respect to using CMB data alone) the constraint on the CDM density parameter (shrinking the 95\% limit by $\sim 40\%$) and on the Hubble parameter $H_0$ (whose 95\% limit improves by 20\%), but it is also able to improve by $\sim 14\%$ the 95\% upper bound on the total neutrino mass.

In conclusion, the clustering ratio appears to be a valuable probe to constrain the parameters of the cosmological model, especially with upcoming large galaxy redshift surveys. Being easy to model and to measure, it provides us with a powerful tool to complement other approaches to galaxy clustering analysis, such as the measurement of the galaxy correlation function or power spectrum. Moreover, we note that, given the simplicity of combining the clustering ratio measured in different surveys, we expect its true constraining power to emerge when it will be measured in a number of different datasets.

\begin{landscape}

\begin{table}
\centering
\begin{tabular}{l *{12}{c}}
\hline
	 & \multicolumn{3}{l}{$\Omega_b h^{2}$} 	 & \multicolumn{3}{l}{$\Omega_c h^{2}$} 	 & \multicolumn{3}{l}{$\tau$} 	 & \multicolumn{3}{l}{$M_\nu$} 	\\ 
 \hline
Pl (fixed $w$)	& 0.02222 	& $ \pm 0.00017 $ 	& $ \pm 0.00033 $ 	& 0.11978 	& $ \pm 0.00147 $ 	& $ \pm 0.00291 $ 	& 0.07851 	& $ \pm 0.01713 $ 	& $ \pm 0.03355 $ 	& 0.16722 	& $ < 0.19150 $ 	& $ < 0.49402 $ 	\\ 
Pl + CR (fixed $w$)	& 0.02222 	& $ \pm 0.00016 $ 	& $ \pm 0.00031 $ 	& 0.11972 	& $ \pm 0.00128 $ 	& $ \pm 0.00255 $ 	& 0.07801 	& $ \pm 0.01744 $ 	& $ \pm 0.03330 $ 	& 0.15795 	& $ < 0.18088 $ 	& $ < 0.47835 $ 	\\ 
Pl	& 0.02222 	& $ \pm 0.00016 $ 	& $ \pm 0.00034 $ 	& 0.11971 	& $ \pm 0.00142 $ 	& $ \pm 0.00281 $ 	& 0.07737 	& $ \pm 0.01793 $ 	& $ \pm 0.03483 $ 	& 0.22153 	& $ < 0.26698 $ 	& $ < 0.60851 $ 	\\ 
Pl + CR	& 0.02216 	& $ \pm 0.00017 $ 	& $ \pm 0.00033 $ 	& 0.12042 	& $ \pm 0.00149 $ 	& $ \pm 0.00290 $ 	& 0.07469 	& $ \pm 0.01754 $ 	& $ \pm 0.03403 $ 	& 0.20304 	& $ < 0.24510 $ 	& $ < 0.53081 $ 	\\ 
Pl + CLens	& 0.02217 	& $ \pm 0.00017 $ 	& $ \pm 0.00035 $ 	& 0.11967 	& $ \pm 0.00153 $ 	& $ \pm 0.00299 $ 	& 0.06927 	& $ \pm 0.01749 $ 	& $ \pm 0.03420 $ 	& 0.32882 	& $ \pm 0.19711 $ 	& $ < 0.67219 $ 	\\ 
Pl + BAO	& 0.02225 	& $ \pm 0.00015 $ 	& $ \pm 0.00030 $ 	& 0.11949 	& $ \pm 0.00134 $ 	& $ \pm 0.00263 $ 	& 0.07728 	& $ \pm 0.01705 $ 	& $ \pm 0.03304 $ 	& 0.11571 	& $ < 0.14235 $ 	& $ < 0.30423 $ 	\\ 
Pl + CLens + CR	& 0.02213 	& $ \pm 0.00016 $ 	& $ \pm 0.00032 $ 	& 0.12026 	& $ \pm 0.00145 $ 	& $ \pm 0.00288 $ 	& 0.06909 	& $ \pm 0.01681 $ 	& $ \pm 0.03237 $ 	& 0.29440 	& $ \pm 0.17524 $ 	& $ < 0.59471 $ 	\\ 
Pl + BAO +CR	& 0.02223 	& $ \pm 0.00015 $ 	& $ \pm 0.00030 $ 	& 0.11965 	& $ \pm 0.00132 $ 	& $ \pm 0.00261 $ 	& 0.07731 	& $ \pm 0.01666 $ 	& $ \pm 0.03147 $ 	& 0.10844 	& $ < 0.13143 $ 	& $ < 0.28339 $ 	\\ 
\hline
 & \multicolumn{3}{l}{$w$} 	 & \multicolumn{3}{l}{$\ln(10^{10}A_s)$} 	 & \multicolumn{3}{l}{$n_s$} 	 & \multicolumn{3}{l}{$H_0$} 	\\ 
\hline 
Pl (fixed $w$)	& -1.00000 	& $ - $ 	& $ - $ 	& 3.09167 	& $ \pm 0.03332 $ 	& $ \pm 0.06510 $ 	& 0.96531 	& $ \pm 0.00478 $ 	& $ \pm 0.00951 $ 	& 66.36205 	& $ ^{+1.93320}_{-0.79827} $ 	& $ \pm 3.14533 $ 	\\ 
Pl + CR (fixed $w$)	& -1.00000 	& $ - $ 	& $ - $ 	& 3.09042 	& $ \pm 0.03375 $ 	& $ \pm 0.06497 $ 	& 0.96550 	& $ \pm 0.00456 $ 	& $ \pm 0.00906 $ 	& 66.47015 	& $ ^{+1.72753}_{-0.68157} $ 	& $ \pm 2.93440 $ 	\\ 
Pl	& -1.68615 	& $ \pm 0.29543 $ 	& $ \pm 0.59285 $ 	& 3.08881 	& $ \pm 0.03490 $ 	& $ \pm 0.06751 $ 	& 0.96500 	& $ \pm 0.00473 $ 	& $ \pm 0.00953 $ 	& 86.91446 	& $ ^{+12.41268}_{-4.70920} $ 	& $ \pm 15.59971 $ 	\\ 
Pl + CR	& -1.25376 	& $ \pm 0.24254 $ 	& $ \pm 0.52000 $ 	& 3.08537 	& $ \pm 0.03368 $ 	& $ \pm 0.06550 $ 	& 0.96389 	& $ \pm 0.00487 $ 	& $ \pm 0.00967 $ 	& 73.04263 	& $ \pm 6.83317 $ 	& $ \pm 14.51362 $ 	\\ 
Pl + CLens	& -1.67628 	& $ \pm 0.35984 $ 	& $ \pm 0.66750 $ 	& 3.07148 	& $ \pm 0.03387 $ 	& $ \pm 0.06578 $ 	& 0.96452 	& $ \pm 0.00502 $ 	& $ \pm 0.00978 $ 	& 84.62450 	& $ ^{+14.64438}_{-5.54025} $ 	& $ \pm 16.85512 $ 	\\ 
Pl + BAO	& -1.05867 	& $ \pm 0.07959 $ 	& $ \pm 0.16354 $ 	& 3.08846 	& $ \pm 0.03305 $ 	& $ \pm 0.06428 $ 	& 0.96612 	& $ \pm 0.00451 $ 	& $ \pm 0.00886 $ 	& 68.60048 	& $ \pm 1.67361 $ 	& $ \pm 3.36935 $ 	\\ 
Pl + CLens + CR	& -1.22298 	& $ \pm 0.23981 $ 	& $ \pm 0.51234 $ 	& 3.07259 	& $ \pm 0.03189 $ 	& $ \pm 0.06176 $ 	& 0.96385 	& $ \pm 0.00493 $ 	& $ \pm 0.00946 $ 	& 71.08454 	& $ \pm 6.31034 $ 	& $ \pm 13.48706 $ 	\\ 
Pl + BAO + CR	& -1.05267 	& $ \pm 0.07749 $ 	& $ \pm 0.15731 $ 	& 3.08903 	& $ \pm 0.03209 $ 	& $ \pm 0.06098 $ 	& 0.96595 	& $ \pm 0.00453 $ 	& $ \pm 0.00888 $ 	& 68.42143 	& $ \pm 1.63637 $ 	& $ \pm 3.26416 $ 	\\ 
\hline 
%
\hline
	 & \multicolumn{3}{l}{$ \Omega_b h2$} 	 & \multicolumn{3}{l}{$ \Omega_c h^2$} 	 & \multicolumn{3}{l}{$ \tau$} 	 & \multicolumn{3}{l}{$ M_\nu$} 	\\ 
 \hline
Pl	(fixed $w$)& 0.02244 	& $ \pm 0.00016 $ 	& $ \pm 0.00034 $ 	& 0.11948 	& $ \pm 0.00157 $ 	& $ \pm 0.00314 $ 	& 0.09332 	& $ \pm 0.01644 $ 	& $ \pm 0.03291 $ 	& 0.20477 	& $ < 0.26441 $ 	& $ < 0.42993 $ 	\\ 
Pl + CR (fixed $w$) & 0.02242 	& $ \pm 0.00015 $ 	& $ \pm 0.00030 $ 	& 0.11950 	& $ \pm 0.00136 $ 	& $ \pm 0.00265 $ 	& 0.09199 	& $ \pm 0.01659 $ 	& $ \pm 0.03229 $ 	& 0.20843 	& $ \pm 0.11909 $ 	& $ < 0.41013 $ 	\\ 
\hline
 & \multicolumn{3}{l}{$ w$} 	 & \multicolumn{3}{l}{$ \ln(10^{10} A_s) $} 	 & \multicolumn{3}{l}{$ n_s$} 	 & \multicolumn{3}{l}{$ H_0$} 	\\ 
\hline 
Pl (fixed $w$) & -1.00000 	& $ - $ 	& $ - $ 	& 3.05893 	& $ \pm 0.03169 $ 	& $ \pm 0.06381 $ 	& 0.95937 	& $ \pm 0.00518 $ 	& $ \pm 0.01015 $ 	& 66.65175 	& $ \pm 1.43657 $ 	& $ \pm 2.67394 $ 	\\ 
Pl + CR (fixed $w$) & -1.00000 	& $ - $ 	& $ - $ 	& 3.05620 	& $ \pm 0.03264 $ 	& $ \pm 0.06422 $ 	& 0.95969 	& $ \pm 0.00458 $ 	& $ \pm 0.00918 $ 	& 66.60347 	& $ \pm 1.23252 $ 	& $ \pm 2.41851 $ 	\\ 
\hline 
%
%
%
\hline
	 & \multicolumn{3}{l}{$\Omega_b h^{2}$} 	 & \multicolumn{3}{l}{$\Omega_c h^{2}$} 	 & \multicolumn{3}{l}{$\tau$} 	 & \multicolumn{3}{l}{$M_\nu$} 	\\ 
 \hline
Pl (fixed $w$)	& 0.02243 	& $ \pm 0.00016 $ 	& $ \pm 0.00031 $ 	& 0.11942 	& $ \pm 0.00146 $ 	& $ \pm 0.00290 $ 	& 0.09335 	& $ \pm 0.01747 $ 	& $ \pm 0.03445 $ 	& 0.20513 	& $ \pm 0.12382 $ 	& $ < 0.43157 $ 	\\ 
Pl + CR (fixed $w$)	& 0.02245 	& $ \pm 0.00013 $ 	& $ \pm 0.00026 $ 	& 0.11926 	& $ \pm 0.00084 $ 	& $ \pm 0.00167 $ 	& 0.09422 	& $ \pm 0.01655 $ 	& $ \pm 0.03298 $ 	& 0.17749 	& $ \pm 0.11133 $ 	& $ < 0.37693 $ 	\\ 
\hline
 & \multicolumn{3}{l}{$w$} 	 & \multicolumn{3}{l}{$\ln(10^{10}A_s)$} 	 & \multicolumn{3}{l}{$n_s$} 	 & \multicolumn{3}{l}{$H_0$} 	\\ 
\hline 
Pl (fixed $w$)	& -1.00000 	& $ - $ 	& $ - $ 	& 3.05859 	& $ \pm 0.03378 $ 	& $ \pm 0.06670 $ 	& 0.95983 	& $ \pm 0.00491 $ 	& $ \pm 0.00964 $ 	& 66.66588 	& $ \pm 1.38464 $ 	& $ \pm 2.62998 $ 	\\ 
Pl + CR (fixed $w$)	& -1.00000 	& $ - $ 	& $ - $ 	& 3.06008 	& $ \pm 0.03262 $ 	& $ \pm 0.06476 $ 	& 0.96046 	& $ \pm 0.00443 $ 	& $ \pm 0.00854 $ 	& 66.98261 	& $ \pm 1.12263 $ 	& $ \pm 2.18365 $ 	\\ 
\hline 
\end{tabular}
\caption{Mean, 68\% and 95\% levels of the marginalised posterior distributions. In the first part of the table the datasets of Planck's CMB temperature and polarization anisotropies, BOSS measurement of the BAO peak, and Planck's CMB lensing signal are used in combination with the clustering ratio measured in the SDSS DR7+12 sample. In the second part of the table bestfits have been fixed to the ones of the fiducial cosmology, errors for Planck are obtained from the publicly available Planck parameter covariance matrix and error for the clustering ratio are the ones measured in SDSS DR7 and DR12. Finally, in the last part of the table, bestfits have been fixed to the ones of the fiducial cosmology, errors for Planck are obtained from the publicly available Planck parameter covariance matrix and errors for the clustering ratio are the ones predicted for a Euclid-like galaxy survey, following the procedure described in the text.}
\label{tab::like_results}
\end{table}

\end{landscape}


\section*{Acknowledgements}
LG, JB, CC, and JD acknowledge financial support from the European Research Council through the Darklight Advanced Research Grant (n. 291521). CC acknowledges the support from the grant MIUR PRIN 2015 ``Cosmology and Fundamental Physics: illuminating the Dark Universe with Euclid''. The DEMNUni simulations were carried out at the Tier-0 IBM BG/Q machine, Fermi, of the Centro Interuniversitario del Nord-Est per il Calcolo Elettronico (CINECA, Bologna, Italy), via the five million cpu-hrs budget provided by the Italian SuperComputing Resource Allocation (ISCRA) to the class-A proposal entitled ``The Dark Energy and Massive-Neutrino Universe''.

\bibliography{Bibliography_all.bib}


\end{document}